# The friendship paradox: Causal evidence of its behavioral consequences[#]

Gary Charness

Francesco Feri

Università di Udine;
Royal Holloway University of London

Matthew O. Jackson

Stanford University; Santa Fe Institute;
Monash University

Miguel A. Meléndez-Jiménez

Universidad de Málaga

Matthias Sutter

Max Planck Institute for Behavioral Economics;
University of Cologne; University of Innsbruck

17 October 2026

**Abstract.** We provide a first causal analysis of the behavioral consequences of the friendship paradox—the fact that people's friends in a network have more connections than average. We find that people's behavior is biased by their network position: they do not best respond to what they should infer the average behavior of the population to be (which determines their payoff), but instead best respond to the average behavior of their friends. In these games of complements, the friendship paradox generates a systematic upward distortion in actions and increases behavioral dispersion. These behavioral distortions persist despite learning opportunities: subjects fail to overcome the bias even when relocated within the network which should help them to understand how average population behavior can differ from local friend behavior.

---

[#] We thank Michael Seebauer for excellent support in programming and running the experimental sessions. This experiment was designed together with Gary Charness from UC Santa Barbara who unfortunately passed away while implementing it. We dedicate this paper to him. We acknowledge financial support from the German Research Foundation (Deutsche Forschungsgemeinschaft, DFG) under Germany's Excellence Strategy (EXC 2126/2-390838866), from MICIU/AEI/10.13039/501100011033 through projects PID2021-127736NB-I00 and PID2025-173969NB-I00, from the Regional Government of Andalusia through project P24-01944, and from Universidad de Málaga through project PRO-SEJ194-G-2023. The experiment was pre-registered on AsPredicted (#215241), https://aspredicted.org/gj46rs.pdf. This study was approved by the Ethics Council of the Max Planck Society within the framework of the Generalized Approval of Experiments Following the Protocol that is Standard in Experimental Economics (approval no. 2018_3 / 2021_36 / 2023_12 / 2024_20).

## 1. Introduction

Social media makes the behavior of a highly visible minority observable to audiences of millions. The fact that these few people impact others' behaviors has even given rise to them being commonly referred to as ``influencers.'' Users encounter displays of consumption, lifestyles, and political engagement that shape their understanding of what others do and what they consider normal or appropriate. Concern about the effects of these platforms has prompted governments to consider restrictions on their use, particularly among young people. Australia, for example, introduced restrictions on social media accounts for users under 16 in December 2025. These debates highlight a broader economic question: how do people form beliefs and make decisions when the individuals they observe are systematically *unrepresentative* of the population?

The prominence of "influencers" makes this question particularly salient. Their large audiences give their behavior exceptional visibility, but visibility need not imply representativeness. If highly connected individuals consume more, ``party'' more, or express more extreme views than others, observing them disproportionately can make atypical behavior appear commonplace. Whether this exposure distorts decisions depends on how observers interpret it: do they realize that highly connected individuals behave systematically differently, or do they treat the people they observe as representative of the wider population? Such misperceptions matter for economic outcomes because beliefs about others' behavior and about what is socially appropriate shape consumption, investment, cooperation, and strategic choices.

The key driver of the bias in perceptions based on the people that are observed in a network is the friendship paradox (Feld, 1991): in *all* networks that have any heterogeneity in connections, an individual's neighbors tend to be more connected than average. Highly connected individuals are therefore disproportionately represented in the social information people receive. The mechanism applies to *all* settings in which people observe others through network connections, from friends and colleagues to online contacts. As pointed out above, this bias becomes extreme when coupled with modern social media that allows individuals to be followed by enormous numbers of others. For example, Alipourfard et al. (2020) showed that on Twitter (now X) 98% of people have fewer followers than the people they follow, and that an average following relationship points to a person who has more than 10 times more than the average number of followers. So, the connections that most people have via their networks (friends, colleagues, neighbors, social media, etc.) over-represent high-degree people; i.e., those with an above average number of network connections. Importantly, if highly connected individuals exhibit more extreme ideological positions and behaviors than the population average, their overrepresentation lead observers to perceive more extreme social norms (Jackson, 2019). For example, Hong and Kim (2016) find that US politicians with more extreme ideological positions have more Twitter followers. [1]

[1] The existence of a positive association between ideological extremeness and overconfidence (Ortoleva and Snowberg, 2015) suggests a potential reinforcing mechanism: if individuals with more extreme views also

First formalized by Feld (1991), the friendship paradox was pointed out as a property of nodes in a network. It was embedded by Jackson (2019) into network games where its consequences for perceived and resulting norms of behavior were studied. Jackson noted that it is not only that friends tend to have more connections than average, but because of their extra connectedness they also tend to behave differently than average, and then studied the implications of that fact. In particular, with strategic complementarities, the marginal benefit someone derives from adopting a behavior (extensively or intensively) – e.g., smoking, investing in human capital, adopting a new technology – increases with the number of others who also adopt the behavior.[2] If agents rely on local sampling to estimate the overall prevalence of a behavior, the friendship paradox means that their friends tend to have higher levels of behavior than the population average, and so agents systematically overestimate activity. For instance, regarding health behavior, Valente et al. (2005) present observational evidence showing that, for middle school students, each additional friendship increases the probability of smoking by 5%, and Tucker et al. (2013) report that each additional friendship increases the probability of having tried alcohol by 6%. Given that more popular students are seen by more others (i.e., due to the friendship paradox and the fact of them having more friends), then other students over-perceive the prevalence of smoking and alcohol consumption compared to the overall population norm. This then feeds back to increase consumption overall as students best respond to their inflated view of the population norm.[3]In this paper, we provide the first causal evidence that (i) the friendship paradox systematically biases people's behaviors in direct accordance with the theory, and (ii) people do not overcome these biases even when they are moved into different positions in the network and hence given the opportunity to learn about their biases. Moreover, we provide this evidence in a very clean setting, in which individuals are informed of the network structure and therefore have fully available the information to correct for the bias.

Various theoretical proofs for variations on the friendship paradox itself – that nodes' neighbors are more central than they are – exist (Jackson, 2008, 2019; Higham, 2019; Kumar et al., 2024). But it is important to note that the friendship paradox itself is simply a fact that is true of every network in which some people have more connections than others – and so it exists in pretty much all human networks. The interesting *consequences* of the friendship paradox for biased perceptions and behaviors is what we are interested in testing here. Whether it has consequences for behaviors has never been tested, although it has been hypothesized as a plausible explanation for various behavioral misperceptions and amplified

---

express them with greater confidence, their disproportionate visibility may be compounded by the apparent certainty of their messages.

[2] In games of strategic substitutes, behaviors are still influenced in systematic ways by the bias induced by the friendship paradox. This is analyzed in the appendix of Jackson (2019). We focus on complementarities, where the intuitions are most straightforward.

[3] Related evidence comes from Agranov et al. (2021), who study COVID-19 vaccination intentions and find that social norms and herding outweigh incentives to free-ride on others' vaccination. Their findings highlight why perceptions of the overall population norm can matter for individual health decisions.

norms (e.g., see the discussion in Jackson, 2019). Thus, a causal identification has been missing, despite its potential practical importance for understanding myriad human behaviors.

We present results from a controlled laboratory experiment with 320 subjects in which subjects are randomly assigned to different treatments and also to different network positions within each treatment. Our study not only causally identifies the paradox's consequences for the increased overall actions, as well as how actions vary with different people's degrees, but also addresses the additional following key questions: Can people overcome this bias if they are shown the overall structure of the network and can, in principle, infer that their local sample is non-representative? Can people learn to overcome their biased perceptions if they experience different positions in the network?

In our experiment, we implement two 20-player networks with different degree distributions, such that each subject has between one and four connections to other players. Moreover, subjects are given full knowledge of the network structure. This enables them to infer the population behavior from the observations of their friends, if they realize that their friends are a biased sample. Through this design feature, we make it deliberately more difficult for the friendship paradox to influence behavior, as subjects have all necessary information to correct their observations for network bias, whereas in most real-world networks people would not have that sort of broader network information. Subjects choose numerical actions and their optimal action increases with the average action of all other participants in the network and with their own degree. The latter reflects a degree-taste correlation (Jackson, 2019); i.e., that individuals who experience a higher intrinsic utility from a complementary behavior (a stronger 'taste' for the action) have greater economic incentives to seek out and maintain a larger volume of social interactions or vice versa.

The key to the identification of the friendship paradox's influence on behavior is our treatment variation. In the "local-information'' treatment, subjects observe only the previous actions of their network neighbors. In the "full-information'' treatment, subjects observe the previous actions of all other participants in the network. The comparison between the two treatments isolates the effect of local network sampling. If subjects best respond to the information available in the local treatment as if their neighbors were representative of the population, their actions should be on average higher than in the full-information treatment. If instead subjects understand and correct for the friendship paradox, their choices should be the same across treatments.

In addition to identifying the behavioral consequences of the paradox, we also design our experiment to test whether they can be undone – provided they exist. For this reason, subjects play the game for 30 rounds, grouped in 3 phases of 10 rounds. To test learning, in phase 2 half of the participants are moved to a different position in the same network where they have a different degree, while the other half of the participants remain in the same position. Both aspects of changing positions (for those who move) and having new neighbors (for those who stay) could in principle help subjects realize that their local neighbors are not representative of the whole network, which should promote learning to undo the perception bias. This should not only apply to phase 2, but also to phase 3 in which all participants who changed position in the previous phase are returned to their original positions.

Our results provide clear evidence of the friendship-paradox-induced bias and amplification of behavior. Average actions are higher in the local-information treatment than in the full-information treatment for every degree, phase, and for both of the two different networks. The effect is not transitory: it persists as the game progresses, both within phases and across phases. The difference is statistically significant in most comparisons, especially when aggregating across networks. Moreover, in the local-information treatment actions are more dispersed, and in ways directly predicted by the theory: people with more highly connected neighbors react with higher average actions than people with less connected neighbors. This suggests that subjects react to heterogeneous local information: different neighborhoods generate different observed samples, leading to more dispersed choices. This provides additional evidence that local information affects behavior through the friendship-paradox bias.

A more direct analysis of best-response behavior confirms the mechanism. In the first phase of 10 rounds, subjects in the local-information treatment choose actions close to the best response computed from their neighbors' actions, whereas subjects in the full-information treatment choose actions close to the best response computed from the actions of all other participants. Thus, the evidence shows that subjects do not fully correct for the non-representativeness of their neighborhood sample. Importantly, even across all 30 rounds, subjects seem unable to recognize the bias and correct for it, despite the fact that the bias is costly to them because it reduces their earnings.

We find that experiencing different network positions provides, if any, a negligible, and temporary, corrective effect, as subjects revert to biased benchmarks upon returning to their original positions. Likewise, subjects not moving to different network positions, but having new neighbors in the second phase, also do not learn to correct their bias. This means that the bias is a pervasive phenomenon, even in a network where subjects know the entire composition of it.

Our study also relates to the literature on selection neglect, which examines whether individuals account for the nonrepresentativeness of the information they observe (Esponda and Vespa, 2018; Enke, 2020). We study a pervasive structural source of such selection: the disproportionate visibility of highly connected individuals in social networks. Our contribution goes beyond documenting a failure to correct for biased samples, by identifying how network-generated selection affects behavior in strategic interactions. In games with strategic complementarities, responses to unrepresentative social information inflate behaviors which then further feed back into behavior observed by others, linking individual inference to inflated aggregate outcomes.

The paper also contributes to the experimental literature on social networks and information aggregation and to the literature on implications of the friendship paradox. The latter relates to work that addresses, in more general terms, a local-sample bias.[4] Building on the framework introduced by Jackson (2019), Bjerre-Nielsen and Busch (2022) study local network observation as a problem of statistical inference that leads to misperceptions about others' connectedness. The mechanism studied in their paper is related to the "majority illusion" of Lerman, Yan, and Wu (2016), whereby a rare behavior may appear common locally when it is disproportionately adopted by highly connected individuals. Mamunuru et al. (2025) apply the problem of local-sample bias in social networks to the perception of inequality. They argue that individuals do not evaluate inequality with respect to the entire population, but rather through the subset of people they observe in their social network and show both theoretically and with observational data that network structure can amplify or attenuate perceived inequality. Frick et al. (2022) model a related source of sample-bias. In their model of "assortativity neglect", individuals fail to account for the fact that their local environment is shaped by homophily and other network forces and is therefore not representative of the population. Thus, while the friendship-paradox literature emphasizes distortions generated by degree-weighted sampling, Frick et al. (2022) emphasize that there can be a variety of biases driven by sorting across types. In both cases, local information is misinterpreted as representative of aggregate behavior, with especially important consequences when actions are strategic complements.

Our experiment contributes by testing whether degree-biased local observations generated by the friendship paradox translate into distorted actions in a controlled laboratory network, thus providing the first causal evidence for behavioral consequences of the friendship paradox. Relatedly, Frick et al. (2023) provide a theoretical framework for comparing multi-agent information structures according to how efficiently they eliminate uncertainty and facilitate coordination.[5] Our experiment brings a similar comparison to the laboratory by contrasting a local-information treatment, where observations are distorted by the friendship paradox, with a full-information treatment that removes this source of network-induced misperception.

[4] While the studies discussed as related to ours in the reminder of this section emphasize the biases generated by local network sampling, a less related literature discusses how biased sampling properties can be exploited for policy targeting. That literature uses ideas from the friendship paradox as a means of identifying highly connected individuals in the data, rather than having anything to do with biased perceptions or biased behaviors. In a clever analysis of a flu epidemic, Christakis and Fowler (2010) showed that people named as friends were more likely to catch a flu earlier than random classmates. This is based on an implication of the friendship paradox that people named as friends are likely to have more social interactions than typical students. Kim et al. (2015) conduct a cluster-randomized field experiment in rural Honduras and show that targeting the nominated friends of randomly selected individuals can increase the diffusion of public-health interventions without requiring a complete map of the network. Building on this approach, Airoldi and Christakis (2024) provide large-scale experimental evidence that friendship-nomination targeting can generate wider spillovers in health knowledge and behavior than random targeting, thereby reducing the number of households that must be directly treated. Relatedly, Banerjee et al. (2019) show that community members can identify individuals who are particularly effective at spreading information and that seeding information through these locally nominated "gossips" produces substantially greater diffusion than targeting random individuals or individuals with high social status.

[5] See also Frick et al. (2020). They consider the case where agents observe a random sample of other agents' actions over time, which provides information about the state of the world, and find that small amounts of misperception about the type distribution in the population can prevent information aggregation.

In a related vein, Enke and Zimmermann (2019) offer experimental evidence on correlation neglect in belief formation, finding that subjects tend to form beliefs incorrectly when the complexity of the environment is high enough. That is a distinct and non-network effect, but one that also illustrates challenges in belief formation.

Chaudhuri et al. (2026) provide a theoretical basis for the increased dispersion of behavior, as a function of network position based on heterogeneous perceptions. Our analysis of variation in subjects' behaviors based on their positions provides causal evidence in the direction of their theory. Their analysis is with an unknown network, and so different positions lead to different perceptions of the network. Our evidence provides causal evidence of heterogeneity even beyond that, in that people are swayed by their local position even when they all know the overall network. Alipourfard et al. (2020) show theoretically and using Twitter (X) data that the structure of social feeds can systematically distort perceptions of topic prevalence. More generally, Eom and Jo (2014) demonstrate with observational data that the friendship paradox applies not only to degree, but also to any individual characteristic positively correlated with network centrality, such as productivity, publications, or citations. We extend this work by providing causal evidence of the behavioral implications of the paradox, and by examining also its consequences for the dispersion of actions and whether learning through experiencing different network positions can undo the bias.

Finally, there is also a related literature to our work in psychology. Galesic et al. (2012, 2018) develop the Social Sampling Model, which posits that individuals estimate broader societal traits by retrieving information from their immediate social circles. They argue that apparent cognitive or motivational biases (e.g., false consensus, over/underestimating population averages) are often statistical artifacts caused by sampling from local networks shaped by homophily and structural asymmetries. Our causal evidence supports this view.

The remainder of the paper is organized as follows. Section 2 describes the experimental design. In Section 3 we report our results. Section 4 concludes.

## 2. Experimental Design

### *2.1 The game*

Each player is located in a position in a network with $n$ nodes, with degrees that range from one to four. Every player $i$ chooses an action $x_i \in [0,30]$, with up to two decimal places. The payoff of player $i$ depends on their own action, the average action of all the other players in the network, and a reference value $\theta_i$, determined by their degree $d_i \in \{1, 2, 3, 4\}$, as described by the following formula:

$$u_i(x_i, x_{-i}) = 30 - \left| x_i - \frac{1}{2}(\bar{x}_{-i} + \theta_i) \right|,$$

where

$$\theta_i = 10(d_i - 1)$$

is the reference value and $\bar{x}_{-i}$ is the average action of all other $n-1$ participants. The payoff is maximized when the subject chooses the midpoint between $\bar{x}_{-i}$ and the reference value:

$$x_i^* = \frac{\bar{x}_{-i} + \theta_i}{2}$$

A key feature of the payoff function is that higher-degree types have higher reference values, i.e., higher ideal points. Therefore, in equilibrium, actions are increasing in degree. Some sort of degree-preference dependence is needed in order for the friendship paradox to have consequences.[6]

*2.2 Network environment*

To explore the behavioral consequences of the friendship paradox in the laboratory, we consider two large and irregular networks with $n = 20$ nodes. The overall intention of our network design is to keep a balance by creating networks that cannot instantaneously be captured with all their details (regarding the distribution of links and positions of neighbors), so that they mimic larger (social) networks in this aspect, while at the same time giving subjects a chance to process the full network information and thus correct for the non-representativeness of their local observations induced by the friendship paradox . Finding behavioral distortions under these still manageable conditions suggest that our estimates provide a lower bound of their magnitude in more complex environments. At the same time, the manageable networks with 20 nodes still allow for an implementation in the laboratory (similar in spirit to Charness et al., 2014).

To this end, we consider two connected networks, Network 1 and Network 2 (with degrees that range from one to four), depicted in Figure 1, which generate substantial variation in individual degree, neighbors' average degree, and network position, while keeping the size of the network manageable. These two networks are nested, in the sense that Network 2 results from adding seven links to Network 1. In Network 1, there are eight positions with degree 1, six positions with degree 2, four positions with degree 3, and two positions with degree 4. In Network 2, there are four positions with degree 1, four positions with degree 2, six positions with degree 3, and six positions with degree 4. Hence, the two networks differ in the connectivity and in the distribution of degrees, with Network 2 being denser. Both networks share the same clustering coefficient of zero.

[6] Alternatively, we could have considered complementarities between the agent's own action and the expected action of others weighted by own degree. Jackson (2019) considers a linear-quadratic utility function in which there are complementarities between an agent's action and the (expected) action of each one of the other players they get connected to in the network. In equilibrium, players with a higher degree are expected to be more engaged in the behavior, given the complementarities. He also shows that if the network is endogenized then preference types (i.e., a taste for the action) and degrees become positively correlated. In our setting, we simplify matters by directly assuming that higher degrees have preferences for higher actions.

**Figure 1**: Networks

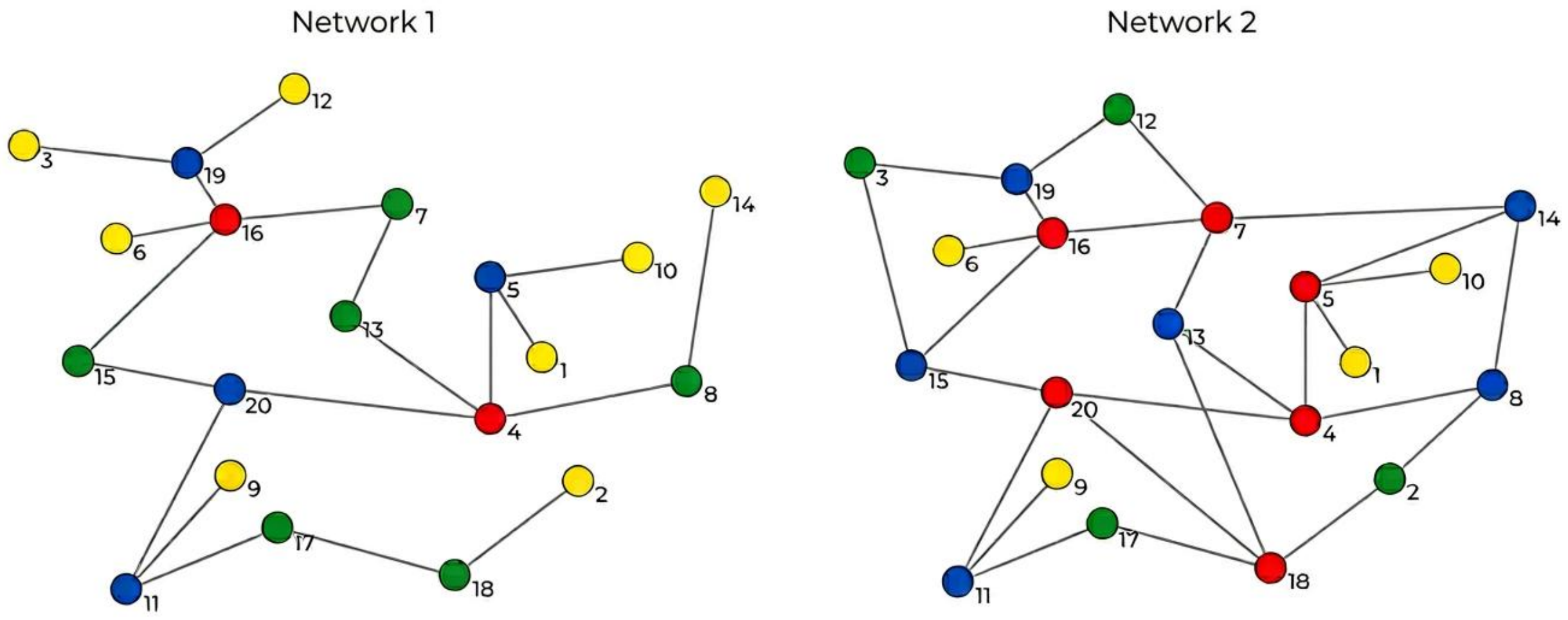

In Tables A1 and A2 in the Appendix, we report the set of degrees of the neighbors for each network position. As we can observe, within each degree there is variability in the sample that each position is linked to, which provides us with additional tools to check for the effects of the friendship paradox. This is so since, if subjects are subjected to the bias, they react to local (rather than global) averages.

*2.3 Timing*

Each experimental session consists of 20 participants. The experiment has three phases, and each phase contains 10 rounds, for a total of 30 rounds. The relatively large number of rounds is intended to allow for extensive learning. At the beginning of the first phase, participants are randomly assigned to network positions. A participant's position remains fixed during the 10 rounds of the first phase. In the second phase, half of the participants remain in the same position, while the other half are moved to a different position with a different degree. In the third phase, all participants return to their original positions from phase 1. This structure allows us to study both the main treatment effect and whether changing positions affects subjects' ability to correct for the friendship-paradox bias, by comparing the behavior in phases 2 and 3 of subjects who change position in phase 2 to those who do not change. The latter may also learn from the positional changes of half of the other subjects as this implies getting new neighbors with potentially different strategies, and this might support learning about whether and how local neighbors are representative of the whole network or not.

*2.4 Experimental design and treatments*

The experiment has two treatments. In treatment LOCAL, at the start of each round subjects observe the previous actions of their direct network neighbors only. For example, in Network 1 the subject in position 15 observes the previous actions of subjects in positions 16 and 20, but not the actions of the remaining participants (see Figure 1). In treatment FULL, at the start of each round subjects observe the previous actions of all other 19 participants in the network.

The first round of each phase is an exception: no previous actions are available, so the corresponding information fields are initially empty. Each round lasts 90 seconds, during which subjects may enter any value as their action and change it as often as they wish. Whenever a subject changes their action, it is immediately visible to those who observe them. Only the last action entered before the 90 seconds run out is the subject's action for that round. Subjects have a payoff calculator and an average calculator available, and following the instructions, a set of control questions is presented to verify that they fully understand the game (see the online appendix for details).

We deliberately designed the experiment so that participants' choices depend on their beliefs about others' average behavior through a simple payoff-maximization problem. Participants know the network structure and the payoff functions, and the calculator allows them to evaluate the payoff implications of any hypothesized average action of the other participants. Given such an average, the unique payoff-maximizing action is its midpoint with the participant's degree-dependent reference value. Thus, if participants understand the game and optimize, their actions directly reflect the average behavior they anticipate: choosing an action amounts to choosing a best response to that anticipated average. The calculator makes this mapping transparent and minimizes the computational demands of implementing it. We therefore chose not to elicit beliefs separately. An additional belief question could direct participants' attention toward population averages and encourage explicit reflection on the representativeness of their neighbors, potentially changing the decision process we seek to study. Our design instead allows us to examine how local information shapes incentivized choices without introducing a separate elicitation task that could itself influence behavior. As we shall see, most participants' actions are rationalized by beliefs that either match the average of their neighbors or the overall average, depending on the subject and the treatment, and so this aspect of the design offers a clear interpretation of behavior.

We run 4 sessions of 20 subjects for each treatment and network. Thus, we have a total of 320 participants, 160 of them randomly assigned to treatment LOCAL (80 randomly for Network 1 and 80 for Network 2) and the other 160 to treatment FULL (80 randomly for Network 1 and 80 for Network 2). The experiment was run in the Decision Lab of the Max Planck Institute for Behavioral Economics in December 2025 and May 2026, using the software oTree (Chen et al. 2016). The experiment lasted, on average, 100 minutes, and the average payoff was €29.34, including a €5 show-up fee. The experimental instructions, translated to English, are reported in the online appendix.

At the end of the experiment we ran a risk test (Charness and Gneezy, 2010), a dictator game, a cognitive reflection test (Thomson and Oppenheimer, 2016) and collected information about gender and age (see the online appendix for details). We can examine how these characteristics correlate with behavior.

In the online appendix we report on a companion prior experiment in which we explore the effects of the friendship paradox in case of a smaller (9-nodes) and symmetric network, only with positions of degrees 1 and 4 that are isomorphic among them. In addition to the network structure, the main differences in design with respect to our main experiment are that we use a linear-quadratic payoff function (which shares the same best-response with our main experiment), and the time structure: We consider ten phases of six rounds each and, although subjects switch positions at the beginning of every phase, they always keep the same degree. Moreover, in addition to treatments with local information about direct neighbors only (called Treatment 1 in the online appendix) and with full information about all network members (called Treatment 2), in the experiment with the smaller network we run two additional treatments: Treatment 3, which is informationally equivalent to Treatment 1, but where the subjects' payoffs depend on their neighbors' (local) average only, rather than on the global average of all other participants; and Treatment 4, which is equivalent to Treatment 1 except for the fact that subjects are not informed of the network structure nor of the payoff functions of other subjects.[7] Treatment 3 is used as a control to study whether subjects are able to properly best respond to the average actions of their neighbors when the incentives are to do so. Since we observe that the distribution of actions follows quite accurately such a (best-response) behavior (see Figure B2 in the online appendix), and there are no relevant differences in how subjects should behave in the small and large network cases (when both information and payoffs are local), we have not run such a treatment as a control in our main experiment.

### *2.5 Predictions*

The main prediction is that actions should be higher in the LOCAL treatment than in the FULL treatment. The reason is the friendship paradox. In LOCAL, subjects observe a sample of neighbors. This sample overrepresents high-degree agents. Since high-degree agents have higher reference values and therefore choose higher actions, the average observed neighbor action tends to be higher than the average action of the population. If subjects best respond to this biased local sample, their actions will be upward biased. The FULL treatment removes this sampling bias because subjects observe all other participants. Therefore, choices in FULL are expected to be closer to the benchmark without the friendship-paradox bias.

[7] As reported in the online appendix, our results suggest that there is small evidence of the friendship paradox bias when networks are small and symmetric, unless there is no information about the network structure (Treatment 4). These findings suggest that the friendship paradox impacts behavior most in complex, asymmetric environments rather than small, symmetric networks.

The theoretical predictions are summarized in Table 1 and detailed in Table A1 in the Appendix. By "equilibrium with bias", we mean the (biased) equilibrium in which subjects best respond to the average action observed among their neighbors. This provides a benchmark for the behavioral distortion generated by the friendship paradox.

In both networks, the predicted actions are systematically higher under the friendship-paradox-biased benchmark than under the no-bias benchmark. This holds for every degree, indicating that local information leads agents to best respond to a biased sample of neighbors and therefore to choose higher actions than they would under the unbiased benchmark. Overall, the theoretical predictions imply a clear treatment difference between LOCAL and FULL: if subjects rely on local information, their actions should be shifted upward relative to the no-bias equilibrium.

**Table 1**: Summary of the equilibrium actions

| | Degree | Equilibrium with bias | | | Equilibrium |
|---|---|---|---|---|---|
| | | average | min | max | without bias |
| Network 1 | 1 | 7.86 | 4.48 | 11.05 | 5.13 |
| | 2 | 12.69 | 8.96 | 15.30 | 10.00 |
| | 3 | 17.12 | 16.42 | 19.08 | 14.87 |
| | 4 | 22.43 | 22.11 | 22.76 | 19.74 |
| Network 2 | 1 | 11.78 | 10.63 | 12.42 | 8.72 |
| | 2 | 16.16 | 15.13 | 16.66 | 13.59 |
| | 3 | 21.23 | 19.37 | 22.99 | 18.46 |
| | 4 | 25.62 | 24.09 | 26.85 | 23.33 |

Regarding the learning, we note the following for those subjects who change position (and degree) in phase 2: (i) by experiencing new positions, they can realize that the observed (local) average is dependent on the position, and therefore, that it is a biased estimate of the global average; (ii) they may be able to observe the behavior of players with degrees that were unobserved in the previous position and (iii) they experience different payoff functions (with different degree-dependent reference values), which may help them to better identify others' incentives. Hence, we predict that changing position (and degree) in phase 2 helps subjects to reduce the bias. Specifically, we hypothesize that the behavior in phases 2 and 3 of subjects who change position should be closer to the unbiased equilibrium than the behavior of those subjects who do not change, as for those shifting positions it seems more salient that their previous neighbors might not be representative of the whole network. In such a case, the actions of subjects who change position would be shifted downwards as compared to the actions of those who do not change. Yet, even for the latter set of subjects (who do not change their position), some learning may happen as they get new neighbors, which should help them realize that the behavior of their old neighbors was likely not representative of the whole set of players with a particular degree. So, even these subjects with no position changes may benefit from others moving around and, therefore, adapt their behavior in phases 2 and 3 towards the no-bias benchmark.

## 3. Results

### *3.1 Average decisions by treatment and degree*

The main result is that subjects choose higher actions in LOCAL than in FULL. This holds in every phase, for every degree, and in both networks. Table 2 reports average actions by treatment, phase, and degree, pooling the two networks. Across all phases, average actions are systematically higher in LOCAL than in FULL for almost all degrees. This pattern is already visible in phase 1 and remains stable in phases 2 and 3, suggesting that the treatment difference is not driven by a specific phase or by a particular network.[8] The same pattern emerges when focusing on the last five rounds of each phase, where learning and adjustment should play a stronger role. Average actions remain higher in LOCAL than in FULL, indicating that the difference between treatments persists as subjects gain experience within the game.

**Table 2**: Average action by phase, treatment and degree

| | | All rounds | | | | Last 5 rounds of each phase | | | |
|---|---|---|---|---|---|---|---|---|---|
| Phase | Treatment | Deg. 1 | Deg. 2 | Deg. 3 | Deg. 4 | Deg. 1 | Deg. 2 | Deg. 3 | Deg. 4 |
| | LOCAL | 8.71 | 13.77 | 18.23 | 23.09 | 8.90 | 14.20 | 18.85 | 24.10 |
| 1 | FULL | 7.46 | 12.05 | 16.55 | 21.25 | 7.10 | 11.69 | 16.56 | 21.15 |
| | | [0.0415] | [0.0074] | [0.0524] | [0.0023] | [0.0249] | [0.0052] | [0.0652] | [0.0002] |
| | LOCAL | 9.12 | 13.47 | 18.40 | 22.59 | 9.06 | 13.82 | 18.73 | 23.16 |
| 2 | FULL | 6.99 | 12.05 | 16.62 | 22.09 | 6.87 | 12.02 | 16.67 | 22.17 |
| | | [0.0415] | [0.0524] | [0.0524] | [0.2209] | [0.0524] | [0.0074] | [0.0974] | [0.1172] |
| | LOCAL | 8.82 | 13.37 | 18.64 | 23.57 | 8.83 | 13.78 | 18.81 | 23.95 |
| 3 | FULL | 7.02 | 11.83 | 16.92 | 22.05 | 6.97 | 11.79 | 17.12 | 22.16 |
| | | [0.0249] | [0.0524] | [0.0803] | [0.0190] | [0.0652] | [0.0190] | [0.0974] | [0.0103] |

[p-values for one-tailed Mann–Whitney test comparing LOCAL and FULL]

One-tailed Mann–Whitney tests[9] comparing LOCAL and FULL support this conclusion. When pooling both networks, most treatment comparisons are significant at conventional levels, both when considering all rounds and when restricting attention to the last five rounds of each phase (see the p-values in square brackets in Table 2). Overall, the evidence points to systematically higher actions under local information than under full information. These results support the prediction that local observation leads to upward-biased choices.

In Figure 2 we depict the evolution of actions over rounds and conditional on degree. It confirms that the treatment difference is not driven only by initial confusion. In both networks, average actions in LOCAL remain above those in FULL across the 30 rounds.[10] The gap is visible for all degrees.

[8] See also Table A3 in the Appendix, which reports the results by networks.

[9] We use one-tailed tests because our preregistered hypothesis was directional: outcomes in the treatment LOCAL were expected to be closer to the (biased) predictions of the friendship paradox than those in treatment FULL. Note that we have 8 independent observations per treatment FULL (4 per network) and 8 independent observations for treatment LOCAL (4 per network).

The persistence of the gap is important. Subjects have repeated opportunities to adjust their actions (within and across rounds), and they receive information about previous actions at the start of each round. Nevertheless, local information continues to generate higher actions. This suggests that the bias is not simply a transitory learning error. Rather, subjects appear to use the observed local sample in a way that does not fully correct for its non-representativeness.

**Figure 2:** Evolution of average actions over rounds by degree and treatment

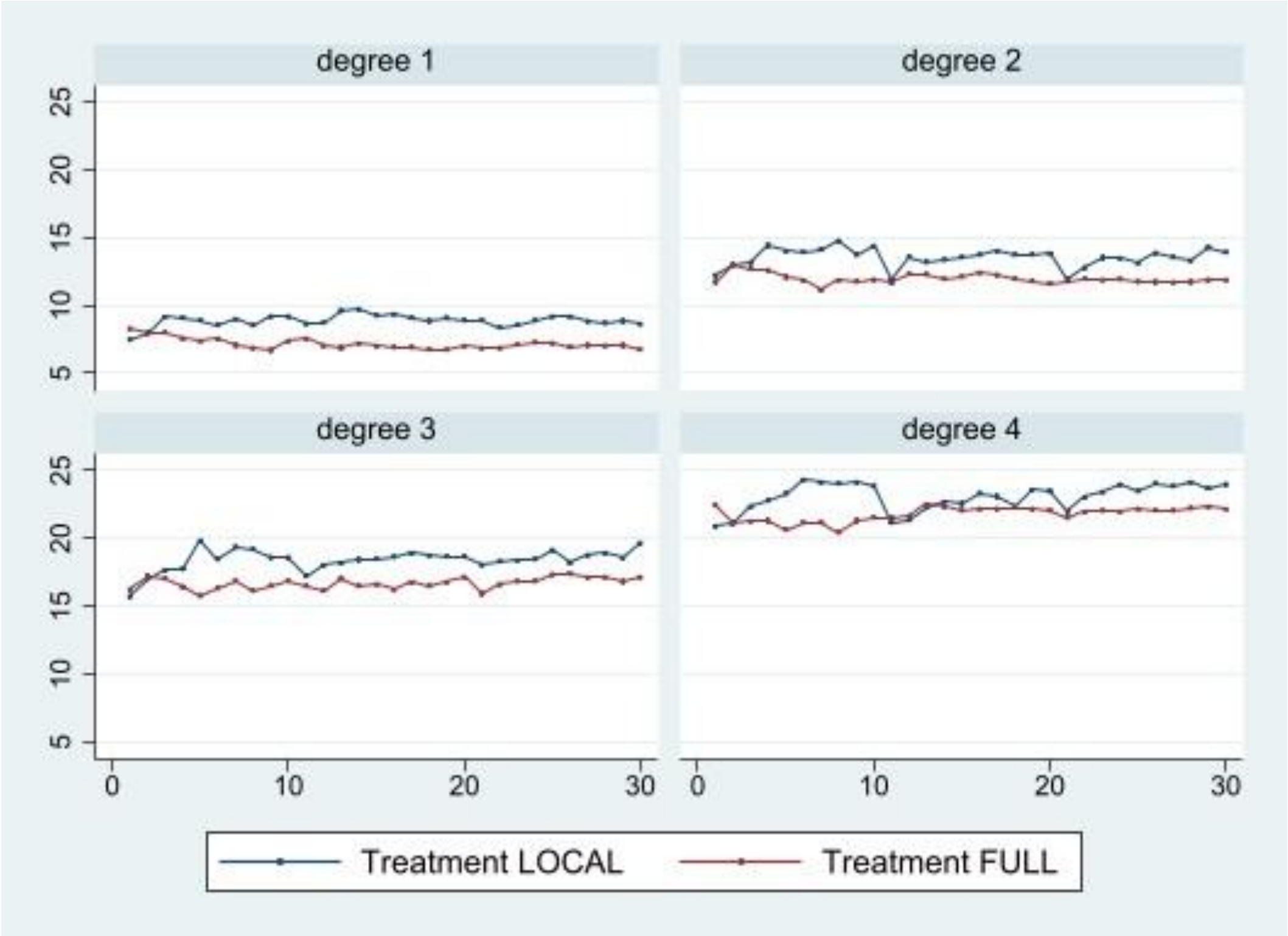

*3.2 Dispersion of decisions*

The local-information treatment increases the dispersion of actions. Conditional on the degree, standard deviations are expected to be zero in FULL, while in LOCAL they are expected to be strictly positive.

Table 3 reports, for each treatment, phase, and degree, the average (across sessions) of the within-session standard deviations of actions, pooling observations from the two networks.[11] Overall, actions are more dispersed in LOCAL than in FULL. This pattern is already visible in phase 1 and becomes even clearer in phases 2 and 3, where LOCAL displays greater dispersion for almost all degrees. The differences between treatments are significant or marginally significant in most cases (see the p-values in square brackets in Table 3), indicating that the local-information treatment generates more heterogeneous choices than the full-information treatment.

[10] The dynamics by network are reported in Figures A1 and A2 in the Appendix.
[11] The results by network are reported in Table A4 in the Appendix.

This higher dispersion in LOCAL is consistent with the interpretation that subjects react to heterogeneous local information. Since individuals observe different neighborhood samples, they are exposed to different signals about others' actions. As a result, local information not only shifts actions upward, as predicted by the friendship-paradox mechanism, but also generates more dispersed behavior. By contrast, in the full-information treatment subjects observe the same relevant set of actions, which reduces heterogeneity in beliefs and decisions.

**Table 3:** Standard deviation of the decisions by phase, treatment and degree

| | | All rounds | | | | Last 5 rounds of each phase | | | |
|---|---|---|---|---|---|---|---|---|---|
| Phase | Treatment | Deg. 1 | Deg. 2 | Deg. 3 | Deg. 4 | Deg. 1 | Deg. 2 | Deg. 3 | Deg. 4 |
| 1 | LOCAL | 4.02 | 4.22 | 3.44 | 2.83 | 3.82 | 3.87 | 3.11 | 2.36 |
| | FULL | 2.61 | 2.26 | 2.07 | 3.30 | 2.31 | 1.80 | 1.94 | 2.75 |
| | | [0.0524] | [0.0325] | [0.0325] | [0.1911] | [0.0415] | [0.0325] | [0.0652] | [0.2209] |
| 2 | LOCAL | 4.44 | 2.97 | 2.92 | 2.32 | 4.17 | 2.93 | 2.63 | 2.12 |
| | FULL | 1.78 | 2.09 | 1.81 | 1.17 | 1.32 | 1.73 | 1.47 | 0.79 |
| | | [0.0015] | [0.0652] | [0.0325] | [0.0415] | [0.0015] | [0.0974] | [0.0249] | [0.0141] |
| 3 | LOCAL | 3.32 | 3.73 | 2.46 | 1.69 | 3.34 | 3.31 | 2.31 | 1.40 |
| | FULL | 1.63 | 1.32 | 0.94 | 1.26 | 1.34 | 1.42 | 0.66 | 1.12 |
| | | [0.0190] | [0.0074] | [0.0103] | [0.1172] | [0.0035] | [0.0141] | [0.0052] | [0.1911] |

[p-values for one-tailed Mann–Whitney test comparing LOCAL and FULL]

### *3.3 Direct evidence on the friendship-paradox mechanism*

To identify the mechanism more directly, we compare subjects' actions with best-response benchmarks. In particular, for each subject and each round 2 to 10 (within each phase) we calculate the following variables:

(a) *avg_local*, which is the average play of the neighbors in round t-1. Then we compute the corresponding "best response" to such an average, *br_local* (i.e., the midpoint between *avg_local* and $\theta_i$). We define *diff_local* as the difference between the subject's decision at round t and *br_local*, *i.e.*, the closer *diff_local* is to 0, the more accurately the subject is (biasedly) "best responding" to the average play of the neighbors in round t-1.

(b) *avg_full*, which is the average play of the overall population (other than the subject) in t-1. We define accordingly *br_full* and *diff_full*.

(c) *avg_forecast*, which is a computed forecast of the average population play if subjects used their neighbors' actions from round t-1 and assigned to each unobserved player the average action of observed neighbors with the same degree.[12] We define accordingly *br_forecast* and *diff_forecast*.

Thus, if *diff_local* is close to 0 in LOCAL, i.e., if subjects best respond to *avg_local*, that indicates that the friendship paradox induces a bias. If *diff_forecast* is close to 0 in LOCAL, i.e., if subjects best respond to *avg_forecast*, then the paradox does not induce a bias. In such a case, the subjects' behavior is consistent with correcting for the degree composition of their local sample. Finally, if *diff_full* is close to 0 in FULL, i.e., if subjects best respond to *avg_full*, then subjects are best responding when they have the whole information.

The evidence from the first phase is shown in Table 4 (see Table A5 in the Appendix for the full table containing information by treatment, phase and type of player). In LOCAL, *diff_local* is close to zero, with an average of -0.12, while *diff_full* and *diff_forecast* are farther from zero, with averages of 1.09 and 0.97. This indicates that subjects behave in LOCAL as if they best respond to the biased local sample. In FULL, the pattern is reversed. The average of *diff_full* is 0.03 and the average of *diff_forecast* is -0.08, both close to zero, while *diff_local* is -1.28. Thus, when subjects observe all other participants in FULL, they best respond to the full-information benchmark rather than to the local-neighbor benchmark.

**Table 4:** Average of *diff_local*, *diff_full* and *diff_forecast* in phase 1

| Treatment | diff_local | diff_full | diff_forecast |
|---|---|---|---|
| LOCAL | -0.12 | 1.09 | 0.97 |
| FULL | -1.28 | 0.03 | -0.08 |

Figure 3 reports kernel-density plots of the three distance measures, providing further support for this interpretation. In LOCAL, the modal behavior is close to the local-information benchmark, whereas in FULL it lies close to the full-information benchmark. The direct best-response analysis therefore confirms that the difference between treatments is driven by how subjects use the network information available to them.

[12] For degrees that are not observed at all, actions are estimated through a linear extrapolation based on the observed relationship between degree and action. If no neighbor with the player's own degree is observed, the player's own action is used for that degree.

**Figure 3:** Kernel density plots of *diff_local*, *diff_full* and *diff_forecast* in phase 1

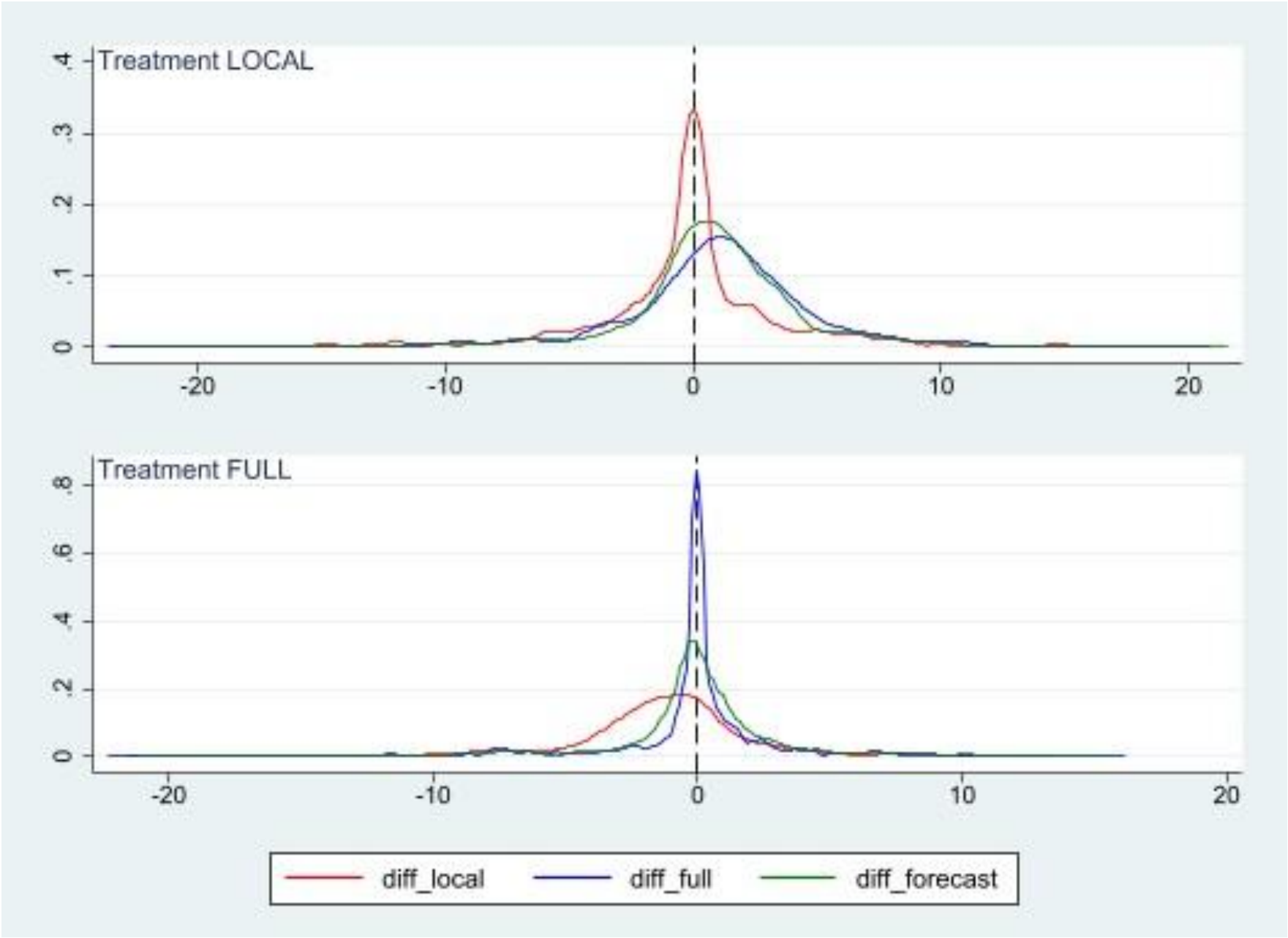

Figure 4 plots the evolution of the mean values of *diff_local, diff_full, and diff_forecast* in each treatment, together with 95% confidence intervals. The absence of any clear time trend provides further evidence that the effect of the friendship paradox is persistent and remains relatively stable across rounds.

**Figure 4:** Dynamics of the averages of *diff_local*, *diff_full* and *diff_forecast* in phase 1

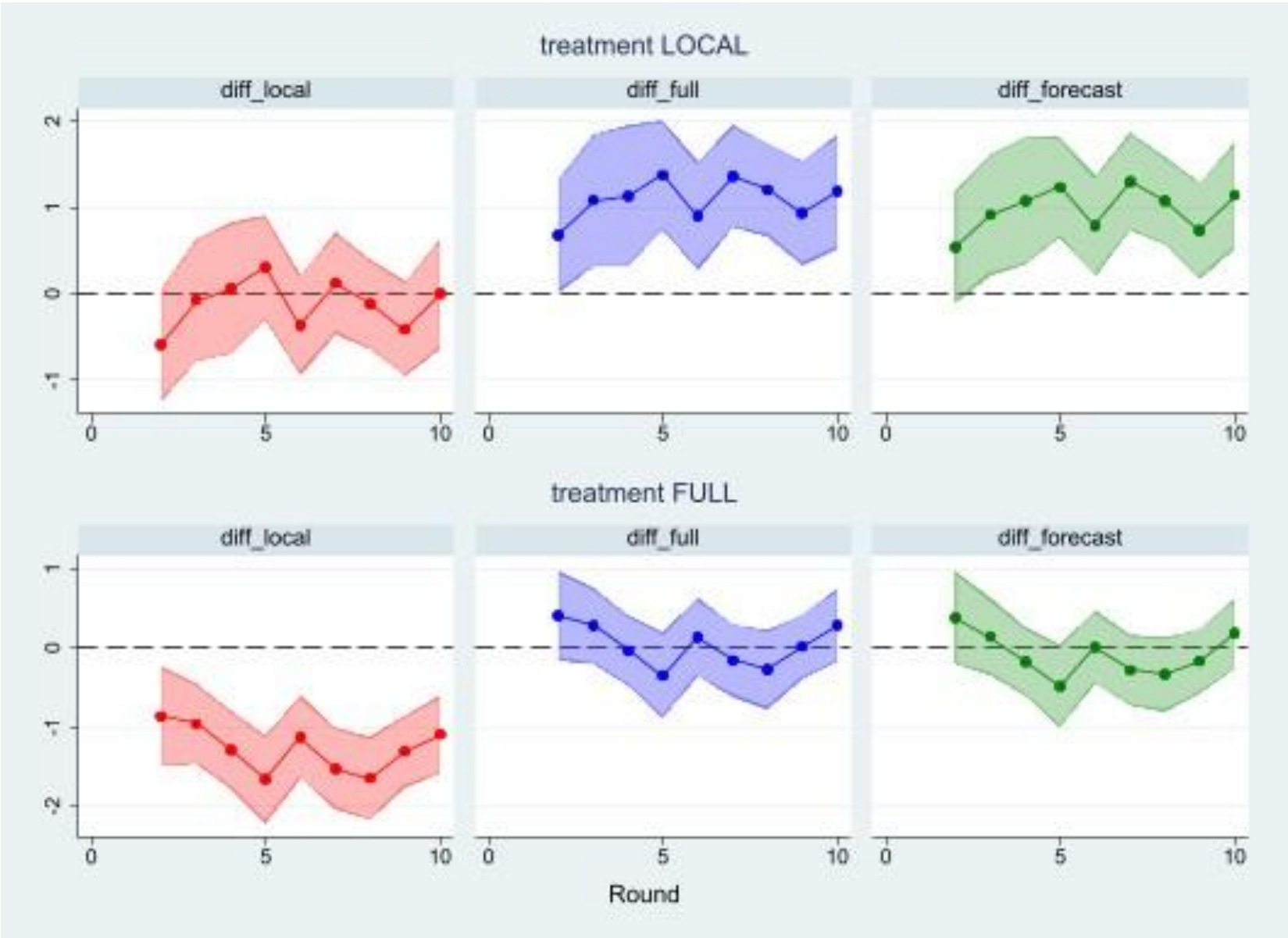

*3.4 Position changes and learning*

The design also allows us to study whether changing position helps subjects correct the friendship-paradox bias, both those who change positions and those who don't. Recall that in phase 2, half of the subjects move to a position with a different degree, while the other half remain in the same position. In phase 3, all subjects return to their original positions from phase 1.

In Table 5 we report average actions for each treatment and degree, separated by type of subject (those who change position and those who do not) and pooling the two networks.[13] Overall, the friendship-paradox bias seems to be robust to changes in network positions, since average actions are systematically higher in LOCAL than in FULL for almost all degrees, even in phases in which some subjects experience (or have already experienced) other network positions.

**Table 5**: Average action by type of subject, phase, treatment and degree

| | | No change in position | | | | Change in position | | | |
|---|---|---|---|---|---|---|---|---|---|
| Phase | Treatment | Deg. 1 | Deg. 2 | Deg. 3 | Deg. 4 | Deg. 1 | Deg. 2 | Deg. 3 | Deg. 4 |
| 1 | LOCAL | 8.01 | 13.43 | 19.32 | 22.27 | 9.40 | 14.12 | 17.14 | 23.92 |
| | FULL | 7.19 | 12.29 | 16.53 | 22.02 | 7.74 | 11.80 | 16.58 | 20.49 |
| | | [0.1393] | [0.0524] | [0.0009] | [0.4392] | [0.0325] | [0.0524] | [0.3227] | [0.0009] |
| 2 | LOCAL | 8.18 | 13.25 | 19.20 | 22.92 | 10.07 | 13.69 | 17.60 | 22.26 |
| | FULL | 6.99 | 12.19 | 16.92 | 22.52 | 7.00 | 11.90 | 16.33 | 21.66 |
| | | [0.1641] | [0.1641] | [0.0190] | [0.2527] | [0.0249] | [0.0803] | [0.1172] | [0.1641] |
| 3 | LOCAL | 8.36 | 13.30 | 19.10 | 23.41 | 9.28 | 13.44 | 18.18 | 23.72 |
| | FULL | 7.06 | 12.21 | 16.99 | 22.37 | 6.97 | 11.44 | 16.84 | 21.72 |
| | | [0.0974] | [0.2209] | [0.0415] | [0.0803] | [0.0103] | [0.0415] | [0.0974] | [0.0249] |

[p-values for one-tailed Mann–Whitney test comparing LOCAL and FULL]

Next, we examine whether, in LOCAL, subjects who later change position and subjects who do not change position behave similarly in phase 1, before any position change occurs. The average data is reported in Table 6 (see Table A5 in the Appendix for the full table containing information by treatment, phase and type of player, including information on statistical tests).

**Table 6:** Average of *diff_local*, *diff_full* and *diff_forecast* in phase 1 of treatment LOCAL, by type of agent

| Type | *diff_local* | *diff_full* | *diff_forecast* |
|---|---|---|---|
| No Change | -0.28 | 0.90 | 0.80 |
| Change | 0.04 | 1.29 | 1.15 |

[13] The evolution of averages across rounds by network is reported in Figures A3 and A4 in the Appendix.

The data suggest that both types (those ones who later change position and those who do not) behave similarly in phase 1, before any position change occurs. Both types are closest to the local-information best-response benchmark. Thus, both groups initially display behavior consistent with the friendship-paradox bias.

If we examine phase 2, we explore if subjects who move in phase 2 may learn from the new position and move closer to the no-bias benchmark, i.e., whether they start to play closer to a best response to *avg_forecast* than they did in phase 1. However, there is no significant evidence in favor of this conjecture. Even if among movers the average of *diff_forecast* falls from 1.15 in the first phase to 0.88 in the second phase, the difference is not statistically significant (see the discussion on Table A5 in the Appendix). In Figure 5 we plot the (kernel) density functions of *diff_ forecast* in the first and second phases both for the subjects who change positions and for those who do not change. The mode is similar in both phases for those who change position, although there seems to be less dispersion in phase 2. Overall, the data suggest that there is not much learning going on, both for those subjects who move and those who do not.

Third, subjects who move and then return to their original position in phase 3 may play closer to the no-bias benchmark than subjects who never moved. The data do not support this conjecture either. In phase 3, both groups remain close to the local-information benchmark. The average values are *diff_local* = -0.29 and *diff_forecast* = 0.93 for subjects who did not change position, and *diff_local* = -0.19 and *diff_forecast* = 0.98 for subjects who changed position, and the values between those who move and those who do not are not significantly different (see the discussion on Table A5 in the Appendix). If anything, subjects who did not change position are slightly closer to the no-bias benchmark. Thus, the experience of moving position does not appear to generate a correction once subjects return to their original position.

**Figure 5:** Kernel density plots of *diff_forecast* in phases 1 and 2 for those who change position and those who do not change

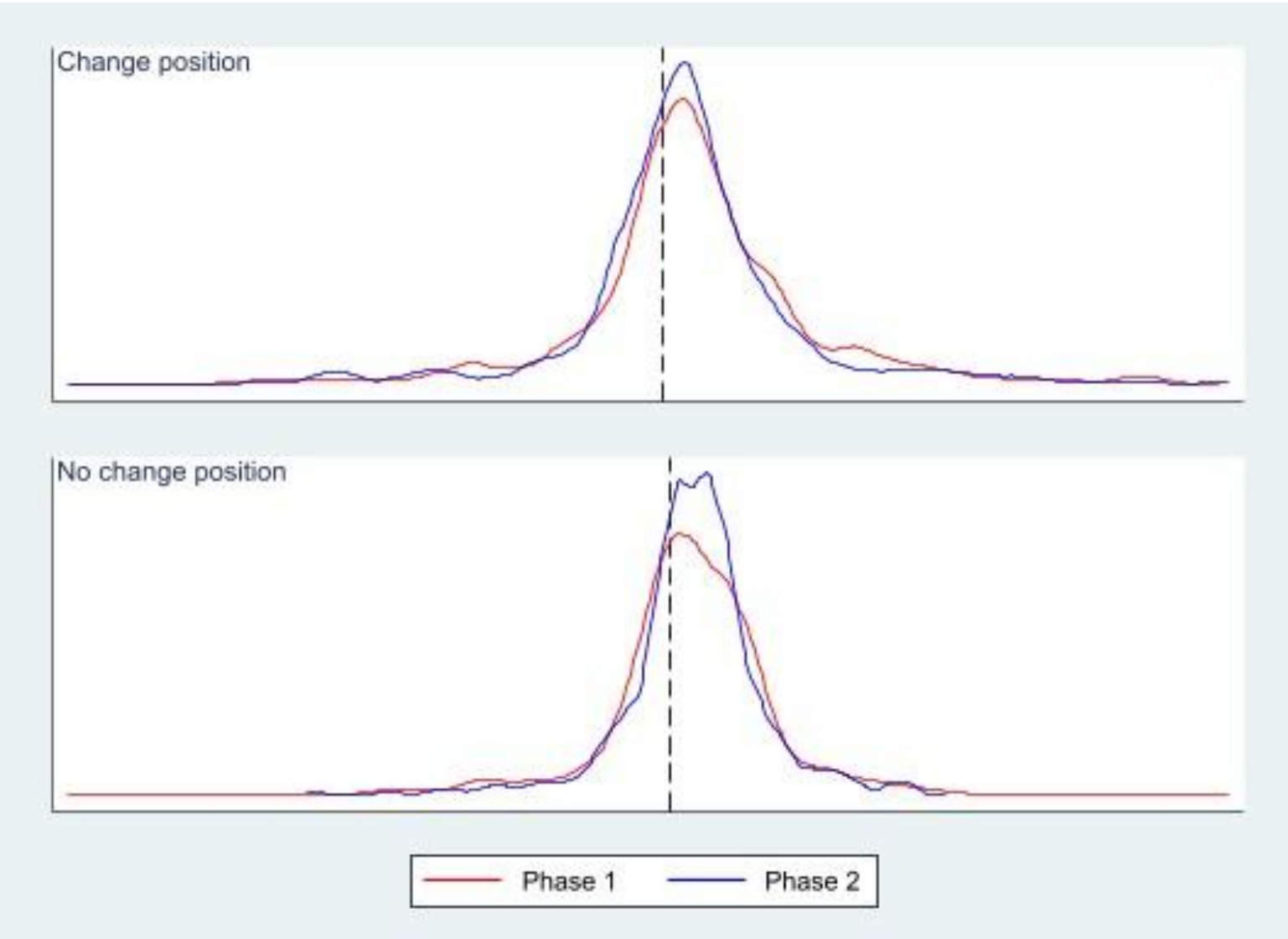

In Figure 6 we plot the (kernel) density functions of *diff_local*, *diff_forecast* and *diff_full* in phase 3, for those subjects who did not change position and for those who did in Treatment LOCAL.[14] The modal play in the upper figure suggests that both types of players are equally close to 0. The modal play in the lower figure suggests that subjects who do not change are closer to 0 (i.e., best respond to *avg_forecast*) than those subjects who do change, although again there is a high dispersion. Thus, modal behavior is not aligned with the conjecture.

**Figure 6:** Kernel density plots of *diff_local*, *diff_forecast* and *diff_full* in phase 3, conditional on changing position or not in Treatment LOCAL

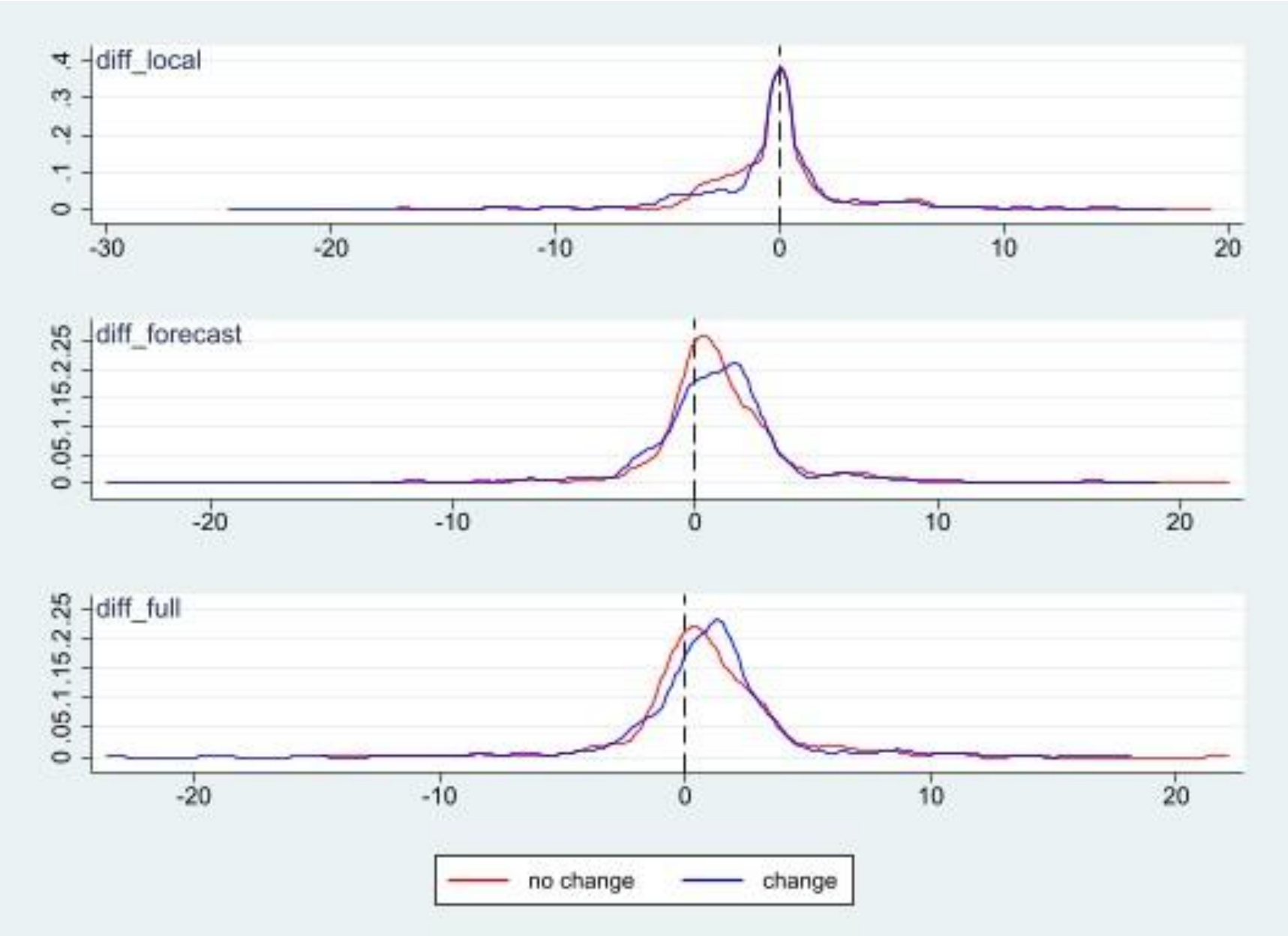

Thus, our results suggest that, overall, position changes provide limited evidence of learning. Moving to a new position may reduce the bias at best temporarily, but the effect does not clearly persist. All of this suggests that the behavioral bias induced by the friendship paradox is a fairly robust phenomenon.

*3.5 Welfare analysis*

In Table 7 we report the average payoff by phase, degree and treatment, pooling the data from both networks. Given the payoff function described in Section 2.1, the (unbiased) equilibrium is efficient, since, given the best response structure, each player attains the maximal attainable payoff (30). We observe that FULL provides an aggregate welfare higher than that of LOCAL, for all degrees and phases. Indeed, we find that almost all differences are statistically significant. This is due to the fact that in FULL subjects have available all the information needed to compute the best response. Thus, in LOCAL, the fact that agents perform sub-optimally, due to the friendship-paradox bias, has an effect on lowering welfare.

[14] The (kernel) density functions for Treatment FULL are reported in Figure A6 in the Appendix.

**Table 7:** Average payoff by phase, treatment and degree

| | | All rounds | | | | Last 5 rounds of each phase | | | |
|---|---|---|---|---|---|---|---|---|---|
| Phase | Treatment | Deg. 1 | Deg. 2 | Deg. 3 | Deg. 4 | Deg. 1 | Deg. 2 | Deg. 3 | Deg. 4 |
| 1 | LOCAL | 26.84 | 26.28 | 27.16 | 27.62 | 27.05 | 26.48 | 27.41 | 28.03 |
| | FULL | 28.07 | 28.32 | 28.61 | 27.64 | 28.40 | 28.77 | 28.75 | 28.03 |
| | | [0.0190] | [0.0141] | [0.0035] | [0.2869] | [0.0190] | [0.0052] | [0.0103] | [0.3227] |
| 2 | LOCAL | 26.72 | 27.39 | 27.56 | 27.77 | 27.01 | 27.31 | 27.71 | 28.03 |
| | FULL | 29.02 | 28.57 | 28.75 | 29.14 | 29.34 | 28.85 | 28.94 | 29.38 |
| | | [0.0001] | [0.0524] | [0.0023] | [0.0074] | [0.0003] | [0.0141] | [0.0052] | [0.0103] |
| 3 | LOCAL | 27.49 | 26.92 | 27.93 | 28.45 | 27.57 | 27.22 | 28.08 | 28.65 |
| | FULL | 29.07 | 29.10 | 29.34 | 28.98 | 29.27 | 29.13 | 29.61 | 29.17 |
| | | [0.0009] | [0.0052] | [0.0023] | [0.0052] | [0.0003] | [0.0052] | [0.0009] | [0.0074] |

[p-values for one-tailed Mann–Whitney test comparing LOCAL and FULL]

### *3.6 Individual behavior*

Next, we investigate the determinants of individual behaviors, using panel data regressions (with standard errors clustered at the session level). In particular, in Table 8 we report on 6 regression models. In models (1) and (4) we use the data from treatment LOCAL, and the dependent variable is the absolute value of *diff_local*. Therefore, we explain the deviations from the "biased" best response (i.e., to the uncorrected local average) when the information is local. In models (2) and (5) we also use the data from treatment LOCAL, but the dependent variable is the absolute value of *diff_forecast* and, thus, we explain the deviations from the best response to the corrected/forecasted average of others when the information is local. Finally, in models (3) and (6) we use the data from treatment FULL and the dependent variable is the absolute value of *diff_full*, hence explaining the deviations from the best response when there is full information.

In models (1)-(3) the explanatory variables are: *Network 2*, a dummy that takes value 1 for observations from Network 2 (and value 0 for Network 1); *Degree*, from 1 to 4; *CRT score*, a variable that accounts for the number of correct answers in the cognitive reflection test (from 0 to 4); *Risk*, which represents the amount invested by the subject in the risky asset in the risk test (from 0 to 2, with higher values indicating more risk tolerance); *Selfishness*, which represents the amount kept by the subject in the dictator game (from 0 to 4); *Age*; and *Round*, from 1 to 30; *Female*, a dummy that takes value 1 if the subject indicated "female" in the gender field of the post-experimental questionnaire (and value 0 otherwise); *Unspecified*, a dummy that takes value 1 if the subject's gender was not identified (as male or female) in the gender field (and value 0 otherwise).

In models (4)-(6) we add as explanatory variables the interactions of the phase of the experiment (from 1 to 3 – *Ph.1*, *Ph.2* and *Ph.3*) and the type of subject (those who change position, *Change*, and those who do not change, *No Ch.*), hence having those subjects who (later) do not change in phase 1 as the baseline.

**Table 8**: Econometric analysis

| | LOCAL | | FULL | LOCAL | | FULL |
|---|---|---|---|---|---|---|
| | (1) | (2) | (3) | (4) | (5) | (6) |
| | \|*diff_local*\| | \|*diff_forecast*\| | \|*diff_full*\| | \|*diff_local*\| | \|*diff_forecast*\| | \|*diff_full*\| |
| *Constant* | 4.846*** | 4.879*** | 2.616** | 4.748*** | 4.742*** | 2.690 |
| | (0.577) | (0.581) | (1.330) | (0.701) | (0.550) | (1.384) |
| *Network 2* | 0.320 | -0.046 | 0.119 | 0.311 | -0.055 | 0.116 |
| | (0.219) | (0.277) | (0.226) | (0.220) | (0.283) | (0.225) |
| *Degree* | -0.263*** | -0.217*** | 0.072 | -0.263*** | -0.217*** | 0.072 |
| | (0.084) | (0.045) | (0.084) | (0.084) | (0.045) | (0.084) |
| *CRT score* | -0.216 | -0.257** | -0.810*** | -0.224 | -0.264** | -0.799*** |
| | (0.146) | (0.126) | (0.175) | (0.149) | (0.128) | (0.184) |
| *Risk* | 0.170 | 0.130 | 0.186 | 0.203 | 0.163 | 0.179 |
| | (0.296) | (0.238) | (0.129) | (0.295) | (0.241) | (0.134) |
| *Female* | 0.147 | 0.212 | 0.390** | 0.086 | 0.151 | 0.414** |
| | (0.469) | (0.366) | (0.186) | (0.451) | (0.338) | (0.174) |
| *Unspecified* | -1.095** | -0.881*** | -0.502 | -0.932 | -0.720 | -0.486 |
| | (0.443) | (0.321) | (0.336) | (0.501) | (0.387) | (0.357) |
| *Selfishness* | -0.464*** | -0.315 | 0.070 | -0.466*** | -0.317 | 0.071 |
| | (0.157) | (0.177) | (0.088) | (0.163) | (0.179) | (0.088) |
| *Age* | -0.010 | -0.006 | 0.022 | -0.011 | -0.006 | 0.023 |
| | (0.011) | (0.007) | (0.024) | (0.010) | (0.006) | (0.024) |
| *Round* | -0.027*** | -0.033*** | -0.045*** | -0.055*** | -0.058*** | -0.062*** |
| | (0.004) | (0.003) | (0.007) | (0.008) | (0.007) | (0.014) |
| *Change×Ph.1* | | | | 0.570 | 0.608 | 0.071 |
| | | | | (0.499) | (0.496) | (0.249) |
| *Change×Ph.2* | | | | 0.953 | 0.844 | 0.147 |
| | | | | (0.527) | (0.501) | (0.287) |
| *Change×Ph.3* | | | | 1.007** | 1.013** | 0.548 |
| | | | | (0.488) | (0.479) | (0.353) |
| *No Ch. × Ph.2* | | | | 0.153 | 0.233 | -0.133 |
| | | | | (0.088) | (0.156) | (0.166) |
| *No Ch. × Ph.3* | | | | 0.795*** | 0.670*** | 0.308 |
| | | | | (0.193) | (0.205) | (0.240) |
| # Obs. | 4312 | 4307 | 4314 | 4312 | 4307 | 4314 |

*** p<0.01, ** p<0.05, * p<0.1

The results in Table 8 suggest that the connectivity of the network (which is larger in network 2) does not significantly affect the ability of subjects to best respond in any model.[15] The subject's degree has a significant effect on the ability to best respond in those models corresponding to treatment LOCAL, suggesting that having more information about others allows subjects to make more precise calculations about average behavior and to properly best respond, but it has no effect on treatment FULL, where subjects have the same information about others' (past) behavior regardless of their degree.

[15] The maximum number of observations is 4,320 per regression. The actual sample sizes are slightly smaller because some participants did not submit a decision within the time limit (5 observations in treatment LOCAL and 6 in treatment FULL). These missing decisions also prevent the construction of the relevant best-response benchmarks for some subsequent observations.

Interestingly, the score in the CRT test significantly predicts the ability of subjects to best respond to the corrected/forecasted average in treatment LOCAL (models (2) and (5)), i.e., to correct for the non-representativeness of their local sample, whereas it has no effect on the "biased best response" (models (1) and (4)). The CRT score also significantly predicts the ability to best respond to the (relevant) average in treatment FULL (models (3) and (6)). These results are consistent with a hypothesis that subjects with a higher cognitive reflection level are better able to mitigate the bias induced by the friendship paradox.

Age and risk considerations do not have an effect on subjects' behavior and, with respect to gender, women have less accurate best responses than men in treatment FULL (models (3) and (6)). Additionally, exploratory observations hint at a potential relationship in treatment LOCAL, where more selfish subjects appeared to align their actions more closely with the biased best response. Regarding the time variable, we observe that over time subjects produce more accurate best responses in all models, although the coefficient is economically very small. In any case, the fact that in treatment LOCAL the time variable significantly affects the ability of subjects to best respond both to the corrected/forecasted average (models (2) and (5)) and also to the "biased average" (models (1) and (4)) suggests some form of heterogeneity: while some subjects may be able to correct for the bias (and learn to do it better over time), others would persist on biased behavior (and produce more accurate "biased best responses" over time).

If we examine the effect of changing positions, our results confirm that there are no differences in phase 1 between subjects who later change positions and subjects who do not change (in any model). Regarding phase 2, in model (5) we observe that it does not significantly affect the accuracy of best responses with respect to phase 1, neither for subjects who change position (p-value 0.43), nor for subjects who remain in the same position (p-value 0.14), so there is no evidence of learning for any of them. Finally, if we compare in model (5) the accuracy of best response in phase 3 of subjects who changed positions to that of subjects who didn't move, we observe that it is not significantly different (p-value 0.27), confirming the results of the previous section.

## 4. Conclusion

We have provided the first causal evidence that the friendship paradox affects human behavior. When subjects observe only their neighbors, their actions are systematically higher than when they observe all other participants in their network. This is exactly the direction predicted by the friendship paradox: local neighborhoods over-represent higher-degree players, who have a preference for choosing higher actions.

The treatment effect is robust. It appears in networks with different connectivity and distributions of degrees, and is observed in all three phases, and for all degrees. It persists when focusing on the last five rounds of each phase, suggesting that repeated interaction and experience do not eliminate the bias.

Our best-response analysis confirms the mechanism. In the local-information treatment (LOCAL), subjects choose actions close to the best response based on neighbors' actions only. In the full-information treatment (FULL), they choose actions close to the best response based on the actions of all other participants. The behavior in LOCAL also varies more across positions, and so has not only a higher average but also greater variance than FULL. This implies that subjects do use the information they observe, but do not sufficiently correct for the fact that local network samples are biased. In our experiment with its 20-player networks, subjects could have corrected for this bias, but they did not, which suggests that such a correction could be even less expected in the typically much larger networks that humans are part of that have much larger degree asymmetries and hence potential for bias and for which they hold correspondingly less overall information.

The analysis of position changes suggests that experience with different network positions has at best very limited corrective power. Subjects who switch to a new position move marginally closer to the no-bias benchmark in the second phase, but this improvement does not persist when they return to their original position. Others moving around also does not influence the behavior of those who keep their network position. In this regard we see little learning that could undo the consequences of the friendship paradox.

The broader implication is that network structure can distort behavior even when incentives are clear and the decision problem is simple. In environments where individuals rely on their social contacts to infer aggregate behavior, popularity, norms, or average choices, the friendship paradox can generate persistent distortions.

In fact, the friendship paradox can also lead to a greater dispersion of actions, as we have also seen in our data. Hence, a structural property of networks – i.e., the fact that in *all* networks that have any heterogeneity in connections, an individual's neighbors tend to be more connected than average – can contribute to a greater variation in behavior. When coupled with degree assortativity, this can also explain why one sees different communities persistently taking different actions.

## Appendix

**Table A1:** Predictions in Network 1

| Degree | Network position | Neighbors' degrees | Equilibrium with bias | Equilibrium without bias |
|---|---|---|---|---|
| 1 | 2 | 2 | 4.48 | 5.13 |
| | 14 | 2 | 6.11 | |
| | 3 | 3 | 8.21 | |
| | 12 | 3 | 8.21 | |
| | 9 | 3 | 8.22 | |
| | 1 | 3 | 8.28 | |
| | 10 | 3 | 8.28 | |
| | 6 | 4 | 11.05 | |
| | Average | | 7.86 | 5.13 |
| 2 | 18 | (1, 2) | 8.96 | 10.00 |
| | 17 | (2, 3) | 11.35 | |
| | 8 | (1, 4) | 12.22 | |
| | 7 | (2, 4) | 14.08 | |
| | 13 | (2, 4) | 14.21 | |
| | 15 | (3, 4) | 15.30 | |
| | Average | | 12.69 | 10.00 |
| 3 | 19 | (1, 1, 4) | 16.42 | 14.87 |
| | 5 | (1, 1, 4) | 16.55 | |
| | 11 | (1, 2, 3) | 16.44 | |
| | 20 | (2, 3, 4) | 19.08 | |
| | Average | | 17.12 | 14.87 |
| 4 | 16 | (1, 2, 2, 3) | 22.11 | 19.74 |
| | 4 | (2, 2, 3, 3) | 22.76 | |
| | Average | | 22.43 | 19.74 |

**Table A2:** Predictions in Network 2

| Degree | Network position | Neighbors' degrees | Equilibrium with bias | Equilibrium without bias |
|---|---|---|---|---|
| 1 | 9 | 3 | 10.63 | 8.72 |
| | 1 | 4 | 12.04 | |
| | 10 | 4 | 12.04 | |
| | 6 | 4 | 12.42 | |
| | Average | | 11.78 | 8.72 |
| 2 | 3 | (3, 3) | 15.13 | 13.59 |
| | 12 | (3, 4) | 16.28 | |
| | 2 | (3, 4) | 16.56 | |
| | 17 | (3, 4) | 16.66 | |
| | Average | | 16.16 | 13.59 |
| 3 | 11 | (1, 2, 4) | 21.26 | 18.46 |
| | 19 | (2, 2, 4) | 19.37 | |
| | 8 | (2, 3, 4) | 20.87 | |
| | 15 | (2, 4, 4) | 21.13 | |
| | 14 | (3, 4, 4) | 21.78 | |
| | 13 | (4, 4, 4) | 22.99 | |
| | Average | | 21.23 | 18.46 |
| 4 | 5 | (1, 1, 3, 4) | 24.09 | 23.33 |
| | 16 | (1, 3, 3, 4) | 24.83 | |
| | 18 | (2, 2, 3, 4) | 25.38 | |
| | 7 | (2, 3, 3, 4) | 25.74 | |
| | 20 | (3, 3, 4, 4) | 26.83 | |
| | 4 | (3, 3, 4, 4) | 26.85 | |
| | Average | | 25.62 | 23.33 |

**Table A3:** Average action by Network, Phase, Treatment and Degree

| | | | All rounds | | | | Last 5 rounds of each phase | | | |
|---|---|---|---|---|---|---|---|---|---|---|
| | Phase | Treatment | Deg. 1 | Deg. 2 | Deg. 3 | Deg. 4 | Deg. 1 | Deg. 2 | Deg. 3 | Deg. 4 |
| Network 1 | 1 | LOCAL | 8.39 | 13.65 | 16.51 | 22.61 | 8.41 | 14.06 | 16.69 | 23.76 |
| | | FULL | 7.26 | 11.73 | 15.47 | 20.24 | 6.76 | 11.07 | 15.33 | 20.31 |
| | | | [0.3429] | [0.1000] | [0.2429] | [0.0571] | [0.1714] | [0.0286] | [0.3429] | [0.0143] |
| | 2 | LOCAL | 8.36 | 13.21 | 16.87 | 21.43 | 8.14 | 13.41 | 17.03 | 22.00 |
| | | FULL | 6.02 | 11.34 | 15.62 | 20.34 | 5.70 | 11.20 | 15.59 | 20.60 |
| | | | [0.0286] | [0.0571] | [0.1714] | [0.1714] | [0.0286] | [0.0571] | [0.1714] | [0.1714] |
| | 3 | LOCAL | 8.12 | 13.33 | 16.99 | 22.23 | 8.02 | 13.39 | 17.06 | 22.29 |
| | | FULL | 6.22 | 11.08 | 15.27 | 20.53 | 6.15 | 11.02 | 15.40 | 20.65 |
| | | | [0.0571] | [0.0571] | [0.0143] | [0.0286] | [0.0286] | [0.0286] | [0.0143] | [0.0286] |
| Network 2 | 1 | LOCAL | 9.34 | 13.96 | 19.37 | 23.26 | 9.89 | 14.39 | 20.30 | 24.21 |
| | | FULL | 7.87 | 12.52 | 17.27 | 21.59 | 7.77 | 12.62 | 17.39 | 21.43 |
| | | | [0.1000] | [0.0571] | [0.0143] | [0.0143] | [0.0571] | [0.0571] | [0.0143] | [0.0143] |
| | 2 | LOCAL | 10.65 | 13.86 | 19.42 | 22.97 | 10.91 | 14.42 | 19.86 | 23.55 |
| | | FULL | 8.95 | 13.12 | 17.28 | 22.67 | 9.19 | 13.25 | 17.40 | 22.70 |
| | | | [0.1000] | [0.2429] | [0.0571] | [0.3429] | [0.1714] | [0.1000] | [0.1714] | [0.1714] |
| | 3 | LOCAL | 10.21 | 13.43 | 19.74 | 24.01 | 10.48 | 14.37 | 19.98 | 24.51 |
| | | FULL | 8.60 | 12.94 | 18.01 | 22.55 | 8.61 | 12.94 | 18.26 | 22.67 |
| | | | [0.0143] | [0.3429] | [0.0571] | [0.1000] | [0.0143] | [0.1000] | [0.1000] | [0.0143] |

[p-values for one-tailed Mann–Whitney test comparing LOCAL and FULL]

**Table A4:** Standard deviation of the decisions by Network, Phase, Treatment and Degree

| | | | All rounds | | | | Last 5 rounds of each phase | | | |
|---|---|---|---|---|---|---|---|---|---|---|
| | Phase | Treatment | Deg. 1 | Deg. 2 | Deg. 3 | Deg. 4 | Deg. 1 | Deg. 2 | Deg. 3 | Deg. 4 |
| Network 1 | 1 | LOCAL | 3.57 | 4.53 | 3.10 | 2.65 | 3.50 | 3.97 | 3.16 | 2.46 |
| | | FULL | 2.65 | 2.34 | 2.24 | 3.04 | 2.21 | 1.89 | 2.18 | 2.10 |
| | | | [0.3429] | [0.1000] | [0.3429] | [0.4429] | [0.1714] | [0.0571] | [0.1714] | [0.5000] |
| | 2 | LOCAL | 4.14 | 3.10 | 2.52 | 1.63 | 3.68 | 3.10 | 2.22 | 1.54 |
| | | FULL | 1.31 | 2.04 | 1.25 | 0.94 | 0.85 | 1.70 | 0.76 | 0.49 |
| | | | [0.0143] | [0.1714] | [0.1000] | [0.1714] | [0.0143] | [0.1714] | [0.0143] | [0.1000] |
| | 3 | LOCAL | 3.75 | 3.04 | 2.20 | 1.31 | 3.63 | 3.00 | 2.18 | 1.11 |
| | | FULL | 2.00 | 1.18 | 0.96 | 0.50 | 1.79 | 1.06 | 0.97 | 0.52 |
| | | | [0.1000] | [0.0571] | [0.1714] | [0.0571] | [0.0571] | [0.0571] | [0.1714] | [0.1714] |
| Network 2 | 1 | LOCAL | 4.48 | 3.90 | 3.78 | 3.00 | 4.13 | 3.78 | 3.07 | 2.26 |
| | | FULL | 2.56 | 2.17 | 1.91 | 3.57 | 2.41 | 1.71 | 1.71 | 3.39 |
| | | | [0.1000] | [0.1714] | [0.0143] | [0.2429] | [0.1714] | [0.2429] | [0.1000] | [0.1714] |
| | 2 | LOCAL | 4.74 | 2.84 | 3.31 | 3.01 | 4.65 | 2.76 | 3.04 | 2.70 |
| | | FULL | 2.24 | 2.14 | 2.37 | 1.39 | 1.79 | 1.76 | 2.17 | 1.10 |
| | | | [0.0571] | [0.1714] | [0.1714] | [0.0571] | [0.0571] | [0.3429] | [0.2429] | [0.1000] |
| | 3 | LOCAL | 2.90 | 4.42 | 2.72 | 2.07 | 3.04 | 3.62 | 2.43 | 1.69 |
| | | FULL | 1.27 | 1.46 | 0.92 | 2.02 | 0.89 | 1.78 | 0.35 | 1.73 |
| | | | [0.0571] | [0.0571] | [0.0571] | [0.4429] | [0.0286] | [0.1000] | [0.0143] | [0.4429] |

[p-values for one-tailed Mann–Whitney test comparing LOCAL and FULL]

**Table A5:** Average *diff_local*, *diff_full* and *diff_forecast* by treatment, phase and type of player

| | | *diff_local* | *diff_full* | *diff_forecast* |
|---|---|---|---|---|
| Treatment LOCAL | | | | |
| Phase 1 | No Change | -0.28 | 0.90 | 0.80 |
| | Change | 0.04 | 1.29 | 1.15 |
| | Overall | -0.12 | 1.09 | 0.97 |
| Phase 2 | No Change | -0.24 | 0.96 | 0.91 |
| | Change | -0.17 | 1.03 | 0.88 |
| | Overall | -0.20 | 0.99 | 0.89 |
| Phase 3 | No Change | -0.29 | 0.94 | 0.93 |
| | Change | -0.19 | 1.13 | 0.98 |
| | Overall | -0.24 | 1.04 | 0.96 |
| Treatment FULL | | | | |
| Phase 1 | No Change | -1.11 | 0.23 | 0.09 |
| | Change | -1.44 | -0.16 | -0.26 |
| | Overall | -1.28 | 0.03 | -0.08 |
| Phase 2 | No Change | -1.05 | 0.32 | 0.27 |
| | Change | -1.52 | -0.13 | -0.17 |
| | Overall | -1.29 | 0.10 | 0.05 |
| Phase 3 | No Change | -1.05 | 0.33 | 0.22 |
| | Change | -1.41 | -0.04 | -0.07 |
| | Overall | -1.23 | 0.15 | 0.07 |

By using within-group tests (signed-rank test, signed test and t-test) we do not find any significant difference (either for *diff_local*, *diff_full* or *diff_forecast*) between any pair of phases, for all treatments and types of players (subjects who change position, subjects who do not change, and on aggregate). Additionally, there is not any significant difference (either for *diff_local*, *diff_full* or *diff_forecast*) between subjects who change position and subjects who do not change, for all treatments and phases, although these evidences are weak given the limited number of clusters.

**Figure A1:** Evolution of average actions over rounds by degree and treatment in Network 1

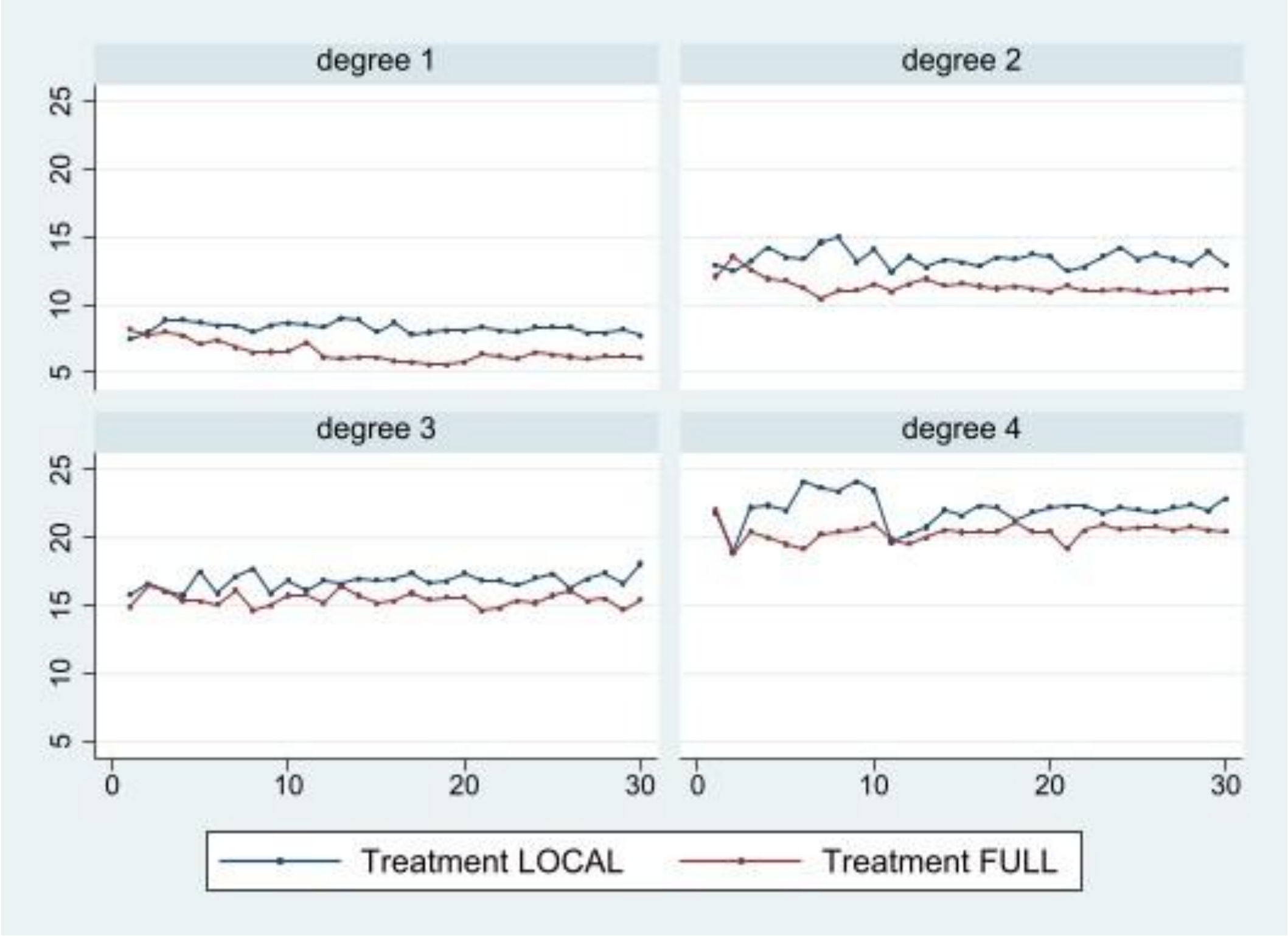

**Figure A2:** Evolution of average actions over rounds by degree and treatment in Network 2

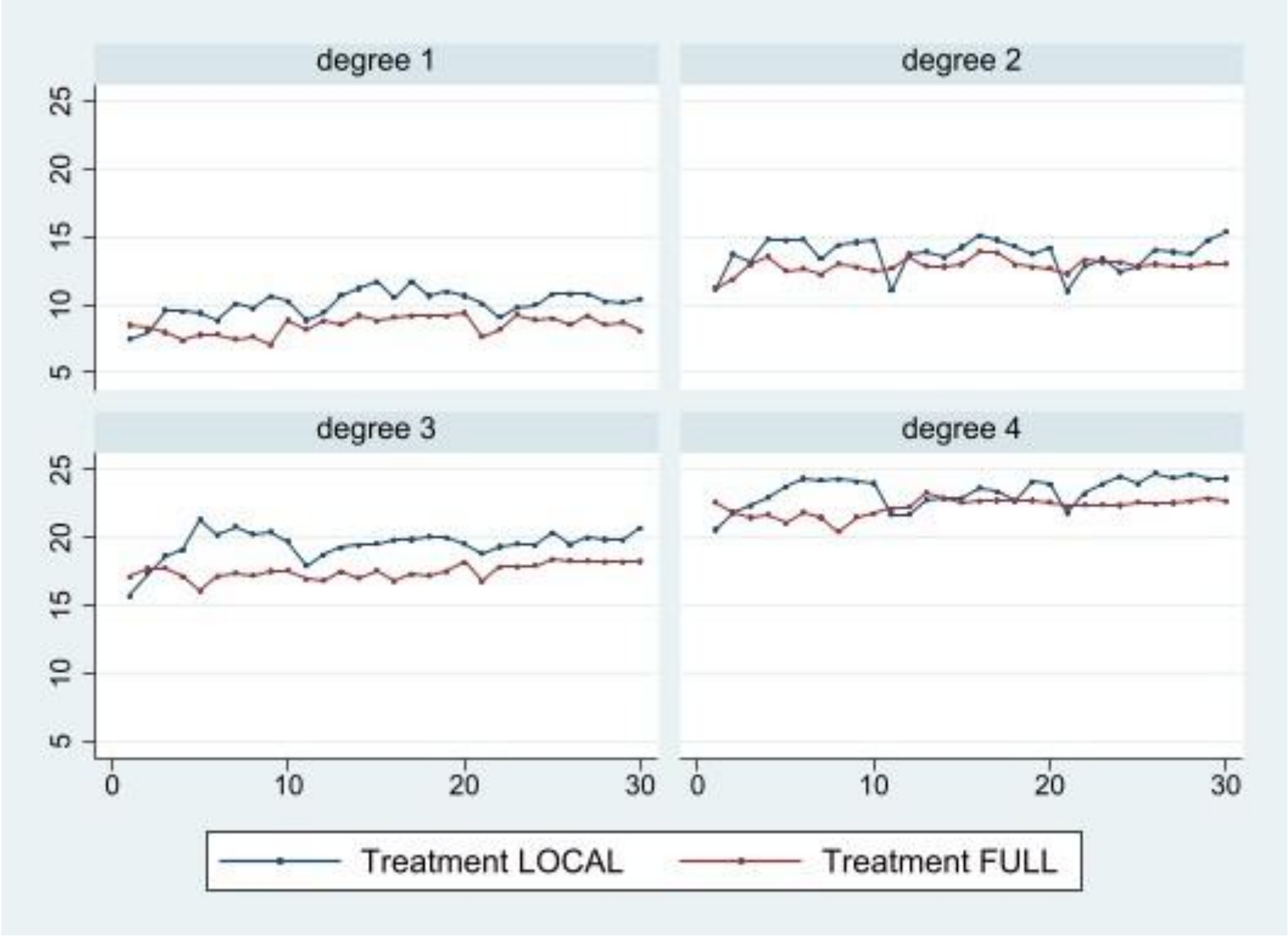

**Figure A3:** Evolution of average actions over rounds by degree and treatment in Network 1 for different types of subjects

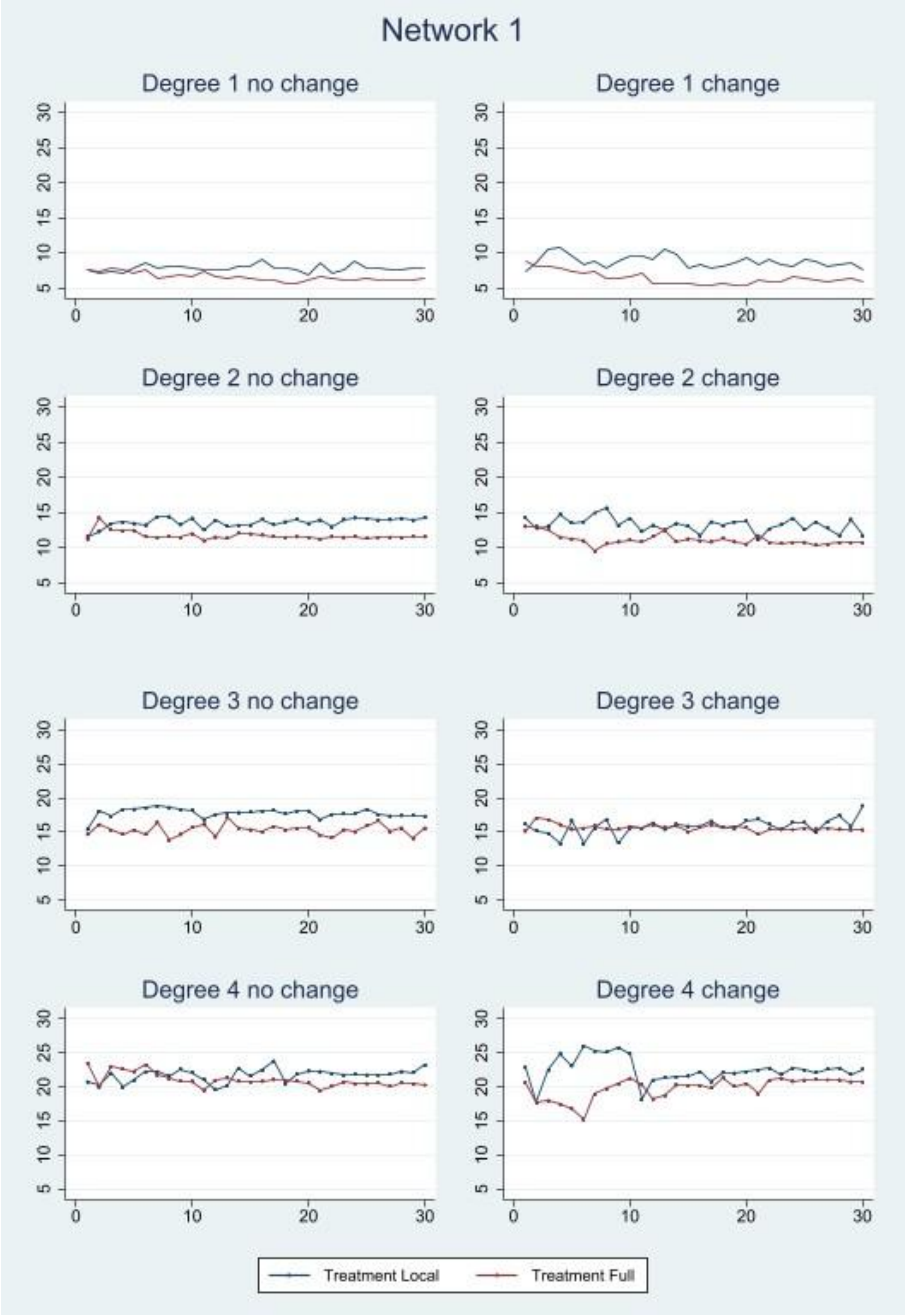

**Figure A4:** Evolution of average actions over rounds by degree and treatment in Network 2 for different types of subjects

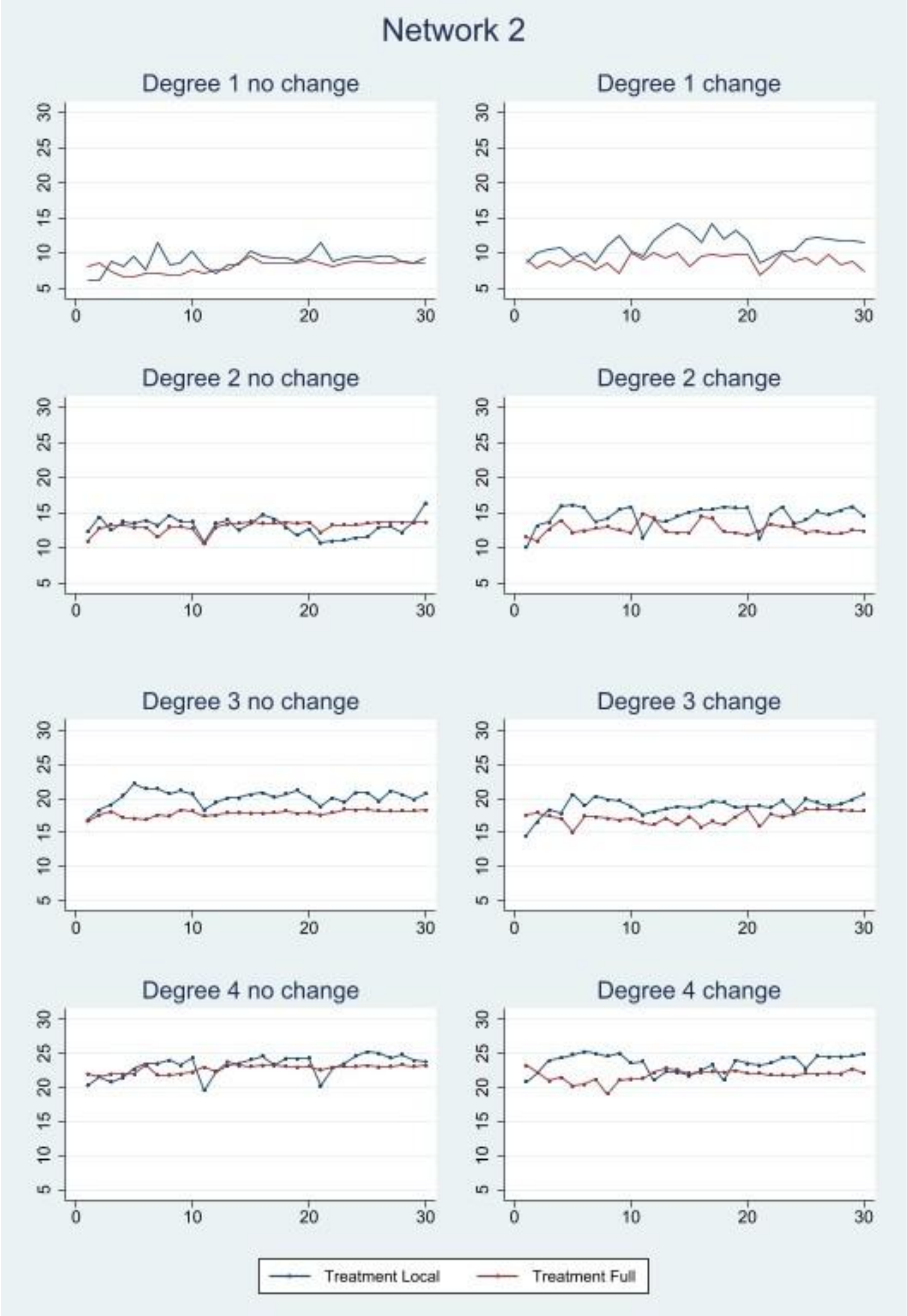

**Figure A5:** Evolution of average actions over rounds by degree and treatment in both networks for different types of subjects

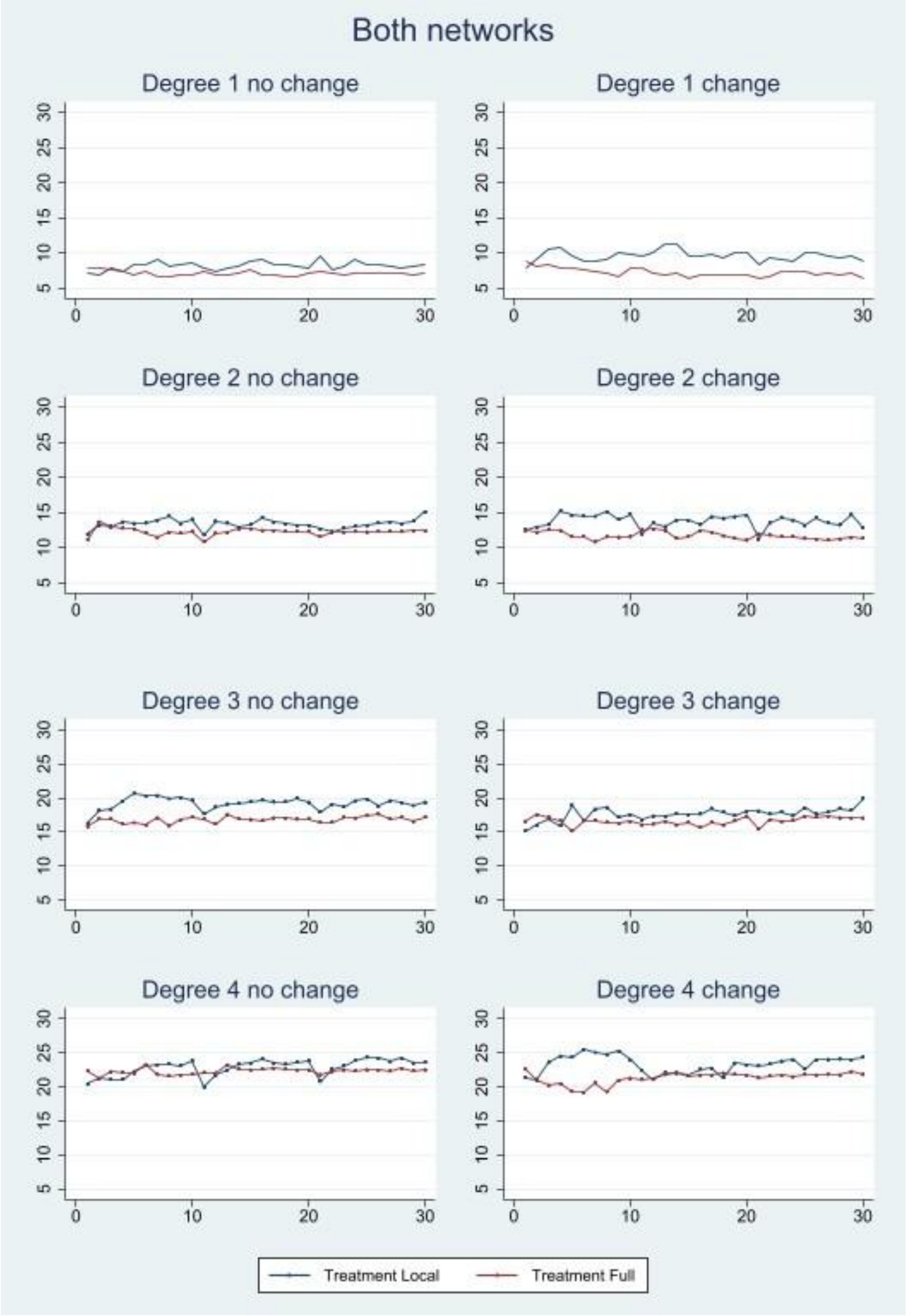

**Figure A6:** Kernel density plots of *diff_local*, *diff_forecast* and *diff_full* in phase 3, conditional on changing position or not in Treatment FULL

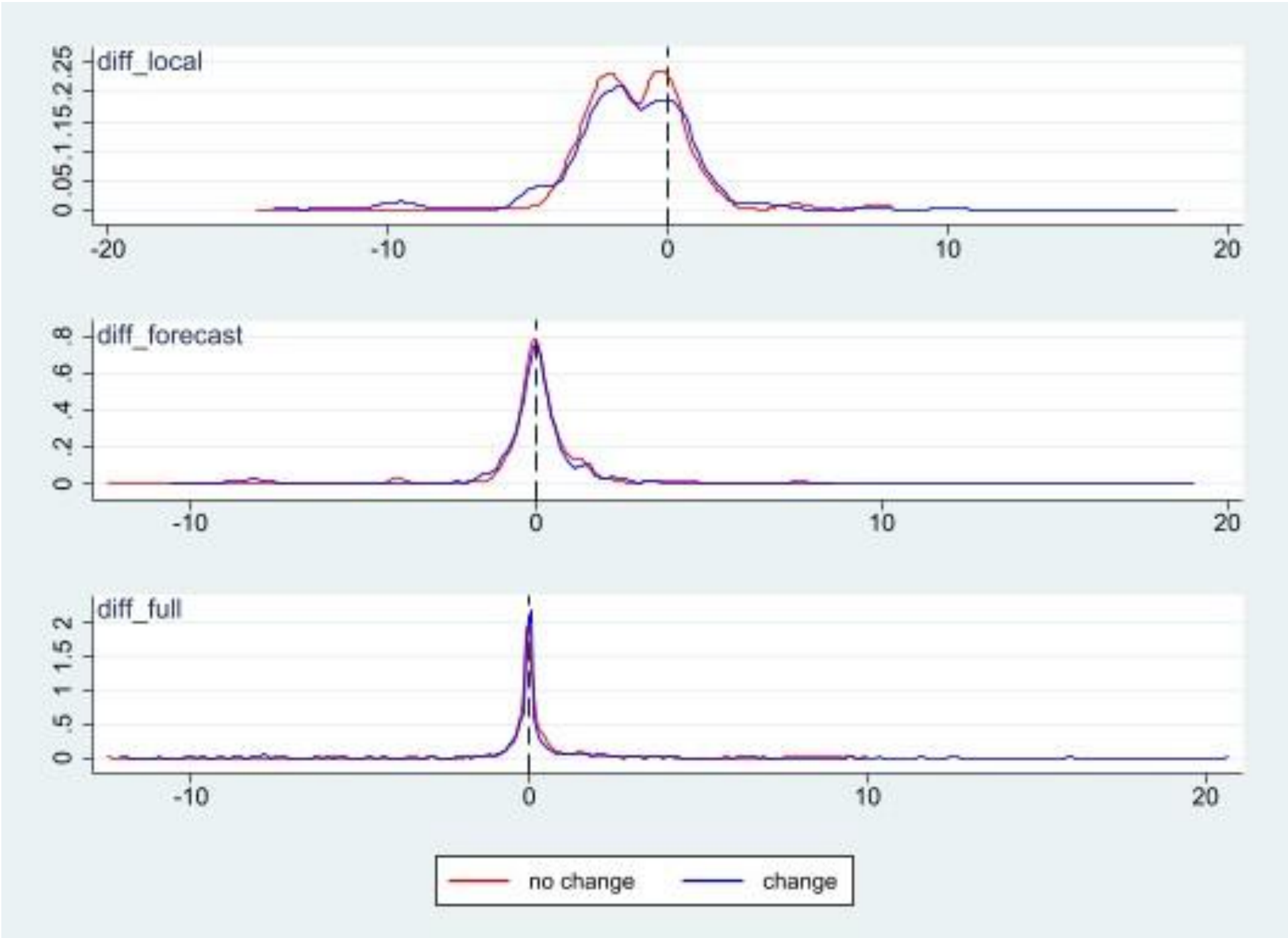

**ONLINE APPENDIX** for
The friendship paradox: Causal evidence of its behavioral consequences

by Charness, Feri, Jackson, Meléndez-Jiménez, Sutter

**– Experimental Instructions**

*Experimental Instructions [LOCAL - Network 1]*

The aim of this experiment is to investigate how people make decisions in certain situations. The instructions are straightforward. If you read them carefully, you can earn a substantial amount of money in cash at the end of the experiment.

The experiment consists of four parts. These are the instructions for the first and most extensive part. You will receive the instructions for the remaining parts at the beginning of each respective part. In the first part of the experiment, your earnings will be calculated in ECU (Experimental Currency Units). Individual payments will remain anonymous, so no participant will learn how much the others have earned. Any communication between participants is strictly prohibited.

1. This part consists of **3 phases**. Each phase is divided into **10 rounds**.
2. Twenty people participate in this experiment. In each phase, each of the 20 participants is assigned to one of the 20 positions in the following network:

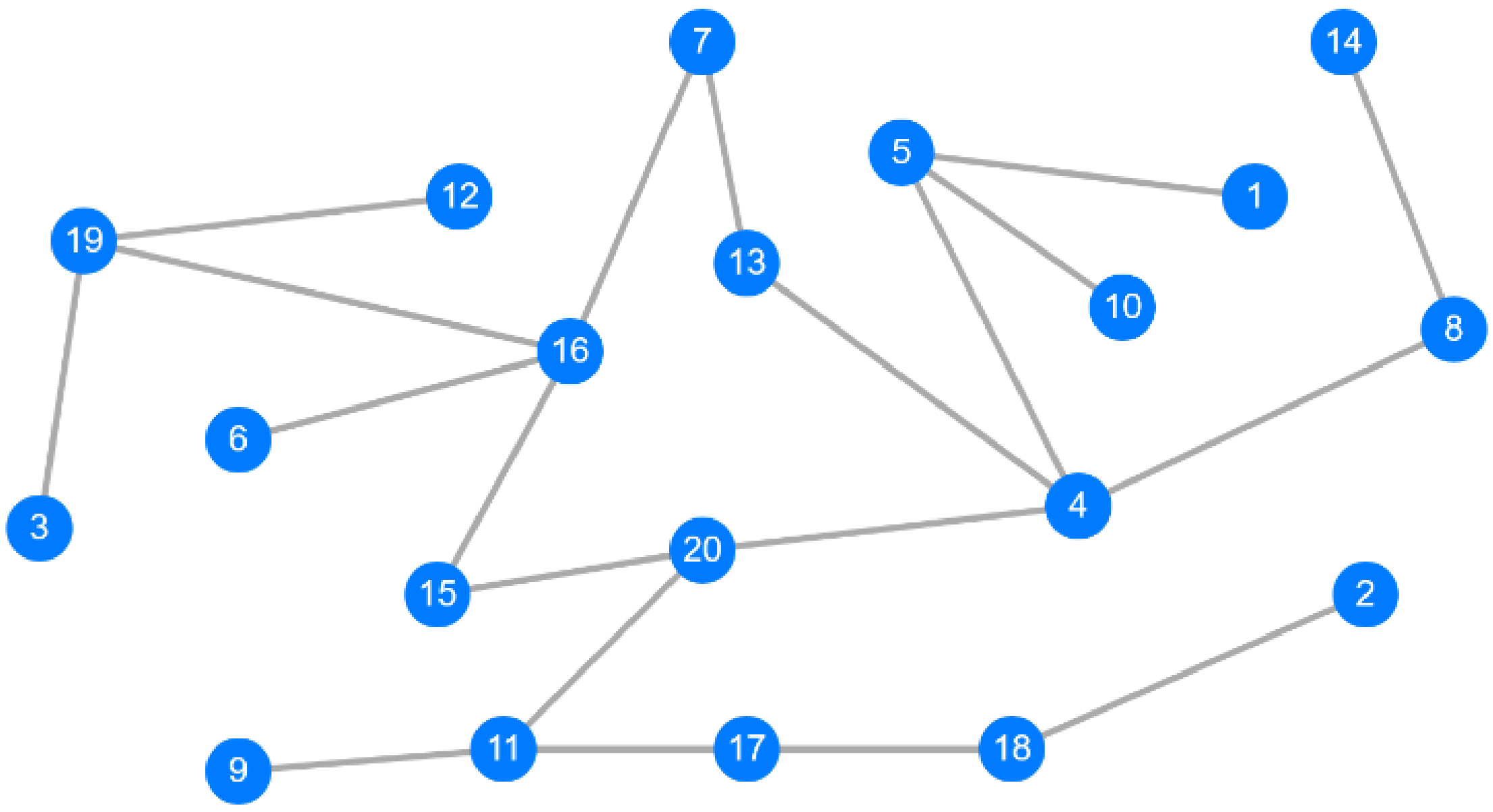

The positions in the network are numbered from 1 to 20. A connection in the network is represented by a line between two positions.

There are eight positions with one connection (positions 1, 2, 3, 6, 9, 10, 12 and 14), six positions with two connections (positions 7, 8, 13, 15, 17 and 18), four positions with three connections (positions 5, 11, 19 and 20), and two positions with four connections (positions 4 and 16). For example, position 13 (with two connections) is connected to positions 4 and 7, but not to any of the other positions.

3. **Assignment of positions**

   At the beginning of each phase of the experiment, you will be assigned to one of the positions in the network. Your assigned position will remain the same throughout all 10 rounds of that phase.

   In the first phase, each participant will be randomly assigned to one of the positions (1 to 20) in the network. In the second phase, half of the participants will remain in the same position, while the other half will be assigned to a different position with a different number of connections from their position in the first phase. In the third phase, all participants will return to their original position from Phase 1.

4. In each round of a phase, every person in the network will be asked to choose a number between 0 and 30, with up to two decimal places. The chosen number is that person's action for the round.

   A person's own chosen action and the actions chosen by the other people in the network determine that person's payoff in the round. We now describe the choice of actions in a round in more detail.

5. Each round lasts 90 seconds, during which every person in the network chooses an action. Please note that you can enter your action in different ways: you can type the desired number directly into the input field and/or use the slider and buttons shown on the screen. You may change your entry as often as you wish before the time limit expires.

   **IMPORTANT:** The number displayed in the "Your Action" field at the end of the 90 seconds is your chosen action for that round.

   At the beginning of a new round, each person will be shown the actions chosen in the previous round by all people who are directly connected to them in the network. Thus, each person sees on the screen the actions chosen in the previous round by the people with whom they share a connection. Each person may change their action at any time during the new round. The new action will then be updated at the beginning of the following round and shown to the people connected to them.

   For example, the person in position 13 will observe the actions of the people in positions 4 and 7 (and vice versa), but not the actions of any of the other people.

   The first round of each phase is an exception: no information about the actions of other people is yet available, and the relevant fields initially display "no information".

6. **Round payoffs**

   Your payoff depends on your own action, the average action of the other 19 people in the network (that is, the sum of the actions of the other 19 people divided by 19), and a reference value determined by your number of connections. If you have 1, 2, 3 or 4 connections, the corresponding reference value is 0, 10, 20 or 30.

- If you are in a **position with 1 connection** (positions 1, 2, 3, 6, 9, 10, 12 or 14), your payoff for the round is calculated as follows:

$$30 - \left| \begin{bmatrix} Your \\ Action \end{bmatrix} - \frac{\begin{bmatrix} Average\ of\ all \\ other\ people's\ actions \end{bmatrix}}{2} \right|$$

- If you are in a **position with 2 connections** (position 7, 8, 13, 15, 17 or 18), your payoff for the round is calculated as follows:

$$30 - \left| \begin{bmatrix} Your \\ Action \end{bmatrix} - \frac{\begin{bmatrix} Average\ of\ all \\ other\ people's\ actions \end{bmatrix} + 10}{2} \right|$$

- If you are in a **position with 3 connections** (position 5, 11, 19 or 20), your payoff for the round is calculated as follows:

$$30 - \left| \begin{bmatrix} Your \\ Action \end{bmatrix} - \frac{\begin{bmatrix} Average\ of\ all \\ other\ people's\ actions \end{bmatrix} + 20}{2} \right|$$

- If you are in a **position with 4 connections** (position 4 or 16), your payoff for the round is calculated as follows:

$$30 - \left| \begin{bmatrix} Your \\ Action \end{bmatrix} - \frac{\begin{bmatrix} Average\ of\ all \\ other\ people's\ actions \end{bmatrix} + 30}{2} \right|$$

In words: your payoff for the round equals 30 minus the distance between your chosen action and the midpoint between the average action of all other people and a reference value determined by your number of connections. If you have 1, 2, 3 or 4 connections, this reference value is 0, 10, 20 or 30, respectively. You can therefore obtain the maximum round payoff of 30 when your chosen action is exactly equal to the midpoint between the average action of all other people and the reference value.

**Please note:** If you have not chosen an action by the time the round expires, your payoff for that round will automatically be set to 0. In the following round, this will be indicated in the relevant field by “no action”.

To allow you to calculate precisely the payoff that may result from your action and the actions of the other people, an “average calculator” will be available in every round. This tool allows you to enter up to 19 numbers and automatically calculates their average. It will be available throughout the full 90 seconds of each round. In addition, the round payoff corresponding to the calculated average and your currently chosen action will be calculated automatically. At the beginning of the experiment, there will be a practice phase in which you can become familiar with the timing of the rounds, the layout of the input screen and the available input methods. The final page of these instructions describes the input screen in detail.

**Please note:** If you have not chosen an action by the time the round expires, you will not be included in the calculation of the average for the other people in the network in that round.

7. **Payment:** At the end of the experiment, exactly one round from Part 1 will be randomly selected for payment. The payoff from the selected round will be converted into cash at the exchange rate

**3 ECU = 2 EURO**

and paid to you in cash after the experiment. In addition, you will receive €5.00 simply for taking part.

**The input screen**

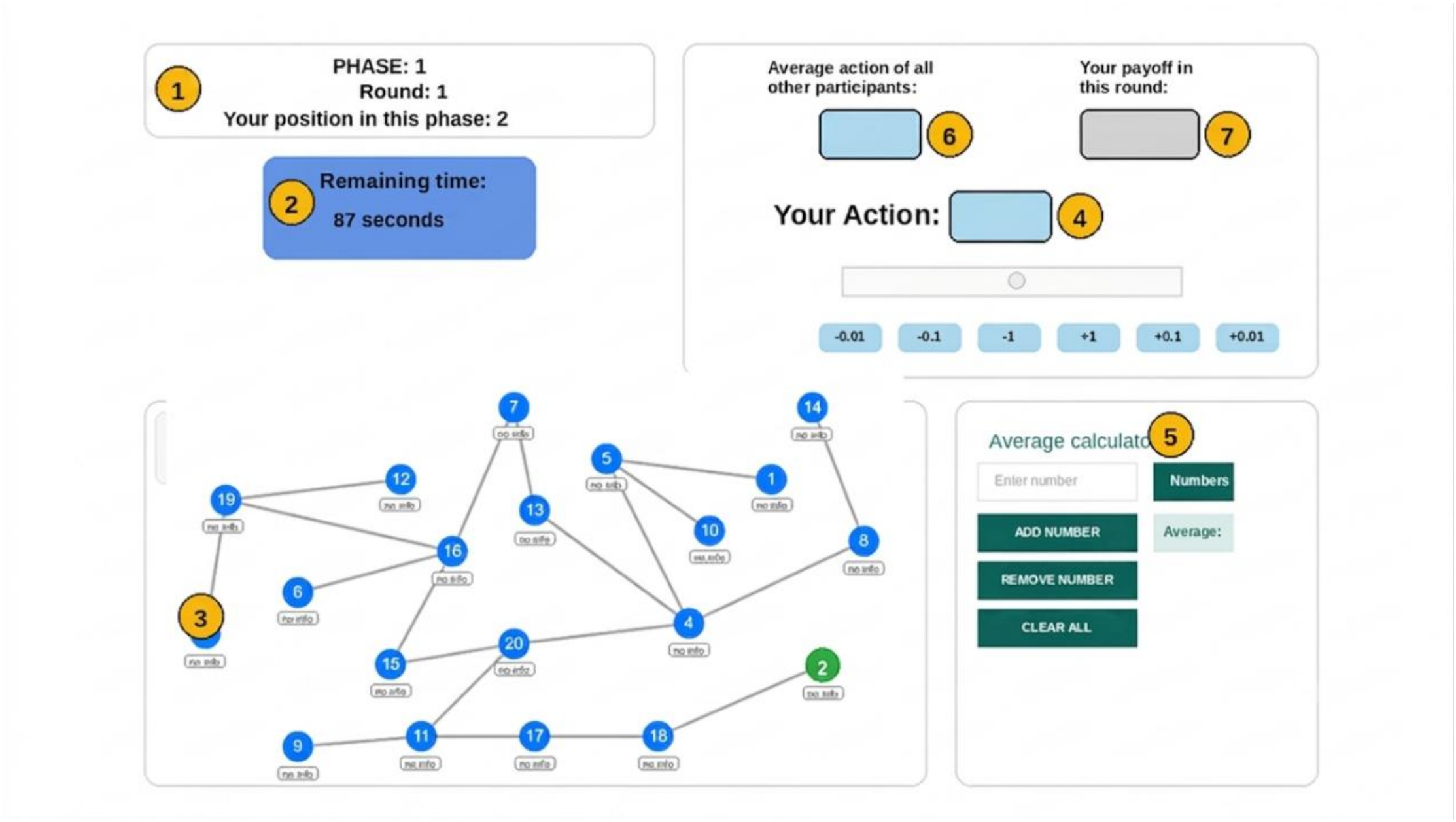

The individual elements of the (still empty) input screen are briefly explained below.

1. The current phase, the current round and your position in this phase.
2. The remaining time in the current round.
3. A representation of the network showing the actions chosen in the previous round of the current phase by the people directly connected to you. You receive no information about any of the other people in the network. Your own position is shown in green. In the first round of each phase, you do not yet have information about anyone.

4. **Your Action:** Here you can enter a number between 0 and 30, with a maximum of two decimal places, as your own action. You can type your action directly into the blue input field. You may also use the slider and/or the buttons below it.
5. **Average calculator:** Here you can enter up to 19 numbers and add them to a list on the right-hand side in order to calculate their average. The currently calculated average is automatically transferred to field (6). To add a number, enter it in the input field and press “Enter” or click “ADD NUMBER”. You can remove individual numbers from the list by selecting them and clicking “REMOVE NUMBER”. Clicking “CLEAR ALL” deletes the entire list.
6. Alternatively, you can directly enter possible average actions of all other people in the network here.
7. **Payoff calculator:** Based on the currently chosen action and the currently entered average action of all other people, the resulting payoff is calculated and displayed automatically here. This is not an input field.

**Please note:** You can actively influence only your own action. The average calculator and payoff calculator are intended solely to help you choose an action. Entering averages does not influence the actions of other people.

At the beginning of the experiment, there will also be control questions about the payoff calculation and a practice phase. This will give you an opportunity to become familiar with the input options and the display of the input screen.

*Experimental Instructions [FULL - Network 1]*

The instructions are the same as in LOCAL, except for point 5, which now reads:

5. Each round lasts 90 seconds, during which every person in the network chooses an action. Please note that you can enter your action in different ways: you can type the desired number directly into the input field and/or use the slider and buttons shown on the screen. You may change your entry as often as you wish before the time limit expires.

   **IMPORTANT:** The number displayed in the “Your Action” field at the end of the 90 seconds is your chosen action for that round.

   At the start of a new round, each participant is shown the actions chosen by all the other 19 participants in the network from the previous round. This means that each person sees on their screen the actions chosen (in the previous round) by the other 19 participants. Each participant can change their action at any time during the new round, and the new action will be updated on the screens of all the other 19 participants at the start of the next round.

   The first round of each phase is an exception: no information about the actions of other people is yet available, and the relevant fields initially display “no information”.

*Control Questions*

[Note: The control questions are the same for both treatments. The only difference is that Question five is correct for T1 and false for T2.]

Please state for each of the following statements whether it is true or false.

These questions are intended to ensure that you have understood the instructions for this experiment well.

| | True | False |
|---|---|---|
| There are 20 people in the network. | o | o |
| Each of these people has exactly two connections to other people. | o | o |
| The positions of all people in the network remain the same throughout the entire experiment. | o | o |
| Within a period, the positions of all people in the network remain the same. | o | o |
| Throughout the entire experiment, you can only see the actions of the people with whom you are directly connected. | o | o |
| Your profit depends only on your decision and the decision of the people directly connected to you. | o | o |
| Every person in the network has the same profit function. | o | o |
| The round profit is calculated differently depending on the number of connections to other people in the network. | o | o |
| In each round, you can change your action as often as you like during the 90 seconds. The action that is in the input field at the end of the 90 seconds will be evaluated. | o | o |
| If no action has been entered at the end of the 90 seconds in a round, the profit for this round is automatically set to 0. | o | o |
| At the end of the experiment, exactly one round from all periods will be randomly selected and paid to you. | o | o |

*Risk Test - Instructions*

You will now receive 2 Euros, which you can invest in a 50:50 lottery. With a 50% probability you win and receive 2.5 times your invested amount, in addition to the portion of the 2 Euros that you did not invest. This means that if you invest your entire 2 Euros, you would receive 5 Euros in the event of a win. With a 50% probability you lose, and receive only the portion of the 2 Euros that you did not invest in the lottery. You can use the slider to set any amount between 0 and 2 Euros in 10-cent intervals.

*Dictator Game - Instructions*

You will be randomly and anonymously grouped with another participant in the experiment. Each of you decides how an amount of money of 4 Euros should be split between the two of you, without knowing the other person's decision. After both decisions have been made, it will be randomly determined whose chosen split will be implemented. The implemented split determines the payoff of both participants. Specify the amount in Euros that you would like to keep for yourself. I would like to keep this amount:

*Cognitive Reflection Test - Instructions*

Below, you will have to answer some questions. For each question you answer correctly, you will receive a payoff of 50 cents.

- Question 1: You are taking part in a race and overtake the person who is in second place. What place are you in then? (Please enter a number as the answer.)
  [Intuitive answer: first; correct answer: second]

- Question 2: A farmer has 15 sheep and all but eight die. How many sheep does the farmer have left?
  [Intuitive answer: 7; correct answer: 8]

- Question 3: Luise's father has five daughters. The first four are named: Lala, Lele, Lili, and Lolo. What is the name of the fifth daughter?
  [Intuitive answer: Lulu; correct answer: Luise]

- Question 4: How many cubic meters of dirt are there in a hole that is 3 m deep x 3 m wide x 3 m long? (Please enter a number as the answer.)
  [Intuitive answer: 27; correct answer: none]

*Questionnaire*

- Please state your age
- Please state your gender

## – Small and symmetric network experiment

### Design

*Network*

We consider groups of 9 subjects, which are fixed for the whole experiment. Subjects are arranged in the 9-nodes network depicted in Figure C1.

**Figure B1:** Small and symmetric network

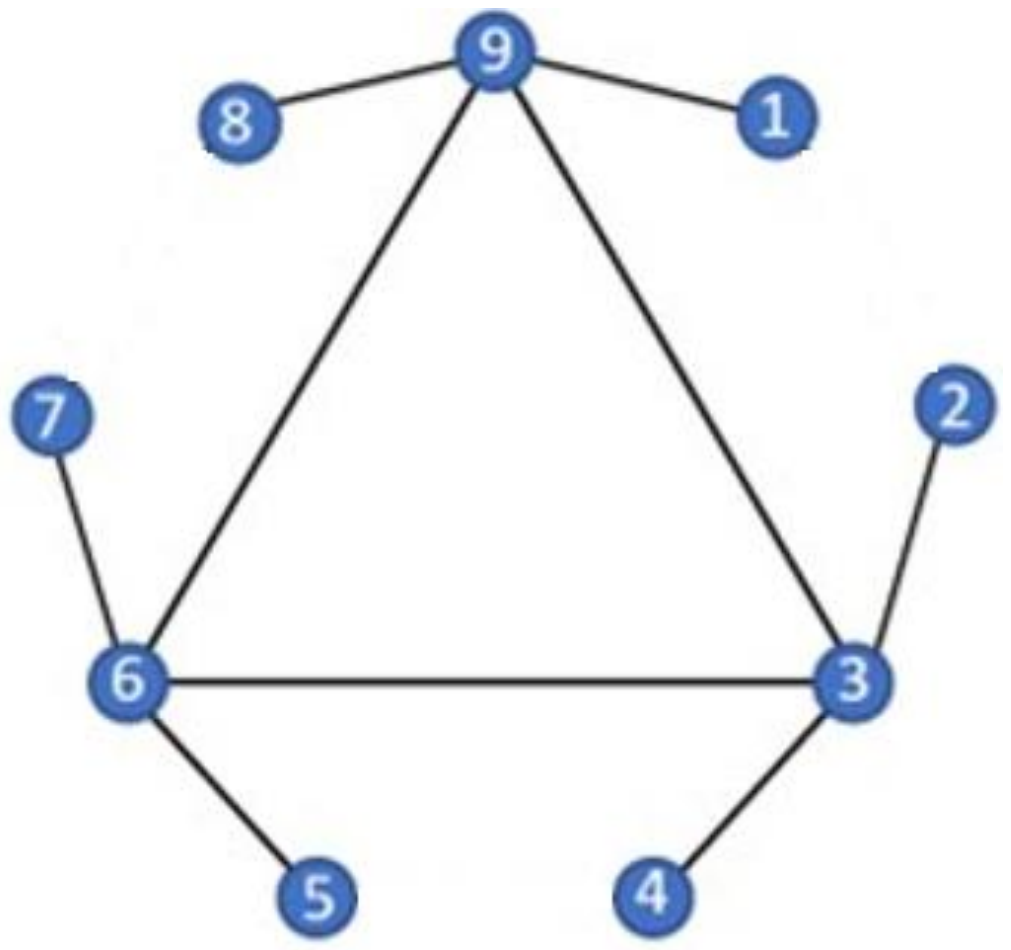

There are 3 subjects with degree 4 and 6 subjects with degree 1 in the network. All positions of the same degree are isomorphic. Each subject has the same degree for the whole experiment. Network positions (within each degree) were randomly re-assigned at the beginning of each phase.

*Time structure*

Subjects play 10 phases; each phase is divided into 6 rounds of 40 seconds. In total, they are playing for 60 rounds. Each phase starts afresh (there is no information about others' behavior in the first round of each phase) and, at each round, subjects receive information about others' behavior in the previous round.

*Treatments and sessions*

We have four treatments (T1, T2, T3 and T4), described in Table B1.

**Table B1:** Treatments for the small and symmetric network experiment

| | | PAYOFFS | |
|---|---|---|---|
| | | Local | Global |
| INFORMATION | Local | T3 | T1 and T4 |
| | Full | -- | T2 |

Treatments mainly differ in two dimensions: information and payoff structure. Regarding information, in T1 and T3 it is local: At each round (except for the first one of a phase) subjects are informed of the actions of their neighbors in the previous round. In T2, there is full information: subjects are informed of the previous actions of all the other 8 players. Regarding payoffs, in T1 and T2 they are global: subjects' payoffs depend on the average action of all other 8 players in the network. In T3 payoffs are local: subjects' payoffs depend on the average action of their neighbors. We note that the combination of full information and local payoffs is empty by design, since it is not of interest.

In particular, the (linear-quadratic) payoff function for a player $i$ with degree $d_i \in \{1,4\}$ is:

$$\pi_i = 1000 - (x_i - \bar{x}_{-i})^2 - (x_i - \theta_i)^2$$

where $\bar{x}_{-i}$ stands for the average action of all other 8 players in the global payoff treatments, and it stands for the average action of player $i$'s neighbors in the local payoff treatment. As in our main experiment, the reference value $\theta_i$ equals 0 for players with degree 1 and 30 for players with degree 4.

Our payoff functions yield the same best response functions as (the linear ones) considered in our main experiment

$$x_i^* = \frac{\bar{x}_{-i} + \theta_i}{2}$$

In the current setup, however, the payoff function is much flatter around the optimum as compared to the linear case, hence yielding weaker incentives to optimize for experimental subjects.

Finally, our last treatment, T4, has the same characteristics as T1 except for the fact that subjects have reduced information: They are neither informed of the network structure nor about the payoff function of other players (they only know their own payoff function).

We have 3 sessions for each treatment, each session consisting of two groups of nine subjects. Thus, we have 6 groups (independent observations) per treatment. In total, we have 216 subjects for the small and symmetric network treatments. The experiment was run in the Decision Lab of the Max Planck Institute for Behavioral Economics. The experiment was pre-registered at AsPredicted (#116606).

**Predictions**

In Table B2 we report the equilibrium predictions, including the prediction with bias in treatments T1 and T4.

**Table B2:** Equilibrium predictions for the small and symmetric network experiment

| Treatment | Degree | Equilibrium with bias | Equilibrium without bias |
|---|---|---|---|
| T1 and T4 | 1 | 12 | 5.29 |
| | 4 | 24 | 19.41 |
| T2 | 1 | -- | 12 |
| | 4 | -- | 24 |
| T3 | 1 | -- | 5.29 |
| | 4 | -- | 19.41 |

**Results**

In Table B3 we report the average action by treatment and degree. We observe that the average actions in T1 are higher than in T2 both for subjects with degree 1 and degree 4, but these differences are small and not significant, suggesting that the bias induced by the paradox is much weaker when we have small networks that are very symmetric (in which it is considerably easier to adjust the observed sample to the overall network configuration). Indeed, when we turn to the comparison between T2 and T4, in which subjects are not informed of the network structure, we observe that the difference becomes significant for subjects with degree 4, yielding more evidence for the behavioral consequences of the paradox.

**Table B3:** Average actions by treatment (T1-T4) and degree

| | T1 | T2 | T3 | T4 |
|---|---|---|---|---|
| Degree 1 | 7.57 | 6.33 | 9.62 | 7.06 |
| | [T1-T2: 0.3939] | | | [T4-T2: 0.5887] |
| | [T1-T3: 0.0260] | | | [T4-T3: 0.0022] |
| | [T1-T4: 0.3939] | | | |
| Degree 4 | 21.01 | 20.70 | 21.06 | 22.56 |
| | [T1-T2: 0.6991] | | | [T4-T2: 0.0260] |
| | [T1-T3: 0.5887] | | | [T4-T3: 0.3939] |
| | [T1-T4: 0.0022] | | | |

[p-value for two-tailed Mann–Whitney test comparing treatments]

When we compare average actions in T3 with those in T1 and T4, we see that, for subjects with degree 1, actions are lower in the treatments with global payoffs, suggesting that they attempt to correct the observed (local) average.

We explore this issue in more detail by comparing subjects' actions with the three different best-response benchmarks introduced in Section 3.3. In Figure B2 we report the kernel-density plots of the differences in each benchmark (*diff_local*, *diff_full* and *diff_forecast*).

**Figure B2:** Kernel density plots of *diff_local*, *diff_full* and *diff_forecast* by treatment

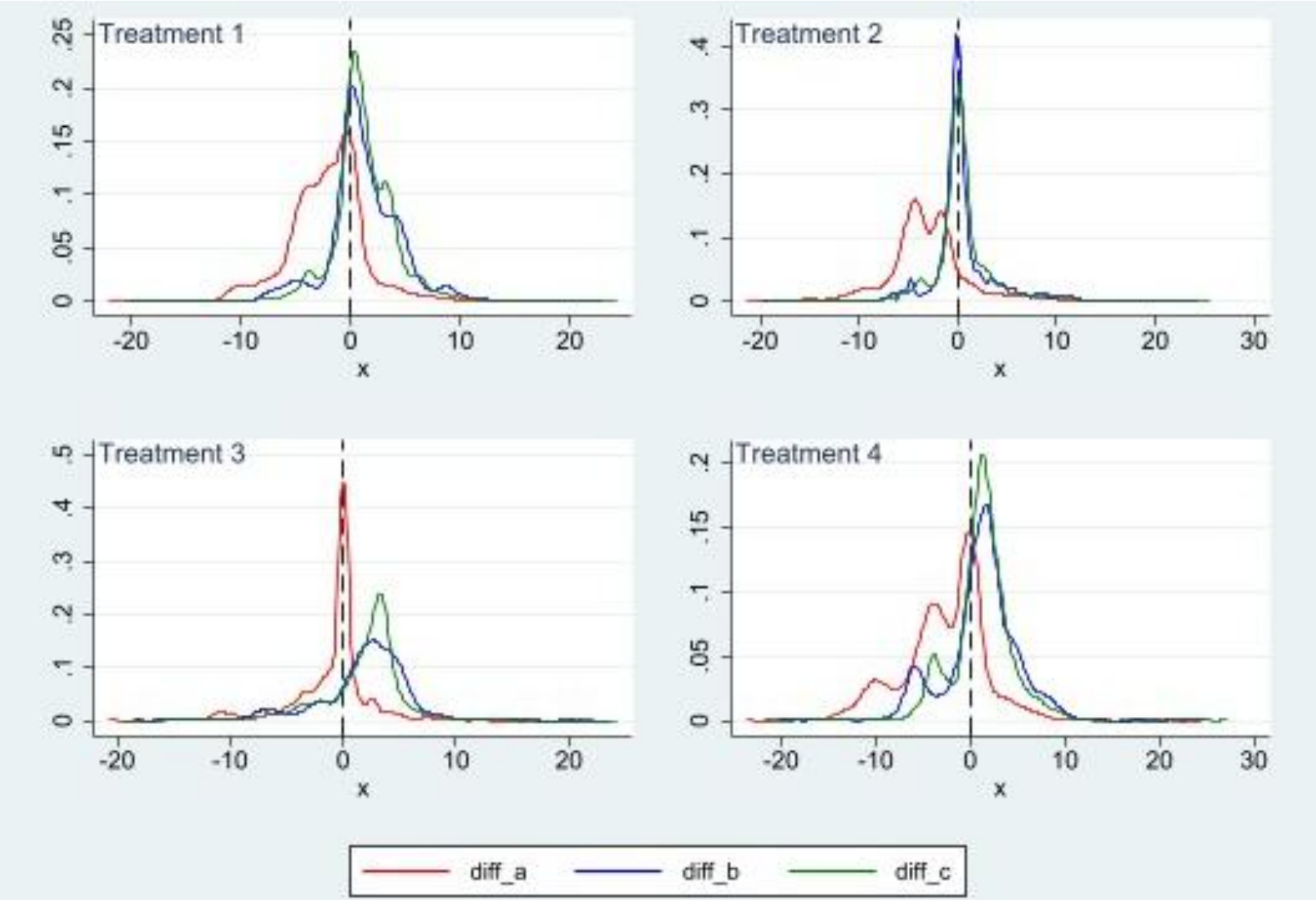

By inspecting Figure B2, we observe that in FULL the modal behavior lies close to the full-information benchmark *avg_full*, and that in Treatment 3 it lies close to the local-information benchmark *avg_local*. This suggests that, in each one of the benchmark scenarios subjects properly respond to the corresponding (global or local) average action.

When we consider Treatments 1 (and 4), where subjects may exhibit biased behavior, and look at the kernel density plot of *diff_local*, we find that, even if the mode is at 0, there is a lot of dispersion, mainly to the left of the distribution, suggesting that there is a significant fraction of subjects trying to correct for the bias. In Treatment 4 there is less dispersion, and the distribution of *diff_local* seems to be more centered at 0 (possibly due to the reduced information about the network subjects have), although there is still a second mode at the left-hand side of the distribution suggesting some attempts for corrections. Indeed, and especially for LOCAL, subjects' behavior seems to be now closer to the no-bias benchmark *avg_forecast*. Thus, unlike our main experiment, these findings suggest that the friendship paradox impacts behavior most in complex, asymmetric environments rather than small, symmetric networks.